\documentclass[prd,twocolumn,unsortedaddress,floatfix]{revtex4}

\usepackage{amssymb}
\usepackage{dcolumn}
\usepackage[dvips]{graphicx}
\usepackage{psfrag}
\usepackage{subfigure}

\usepackage{amsmath,amsfonts}
\usepackage{epsfig}

\begin{document}

\title{Perturbative Wilson loops from unquenched
Monte Carlo simulations \\ at weak couplings}

\author{Kit Yan Wong}
\affiliation{Simon Fraser University, Department of Physics, 
8888 University Drive, Burnaby, British Columbia, Canada V5A 1S6}
\affiliation{Department of Physics and Astronomy, University of Glasgow,
Glasgow, G12 8QQ, United Kingdom}
\thanks{Present address.}

\author{Howard D. Trottier}
\affiliation{Simon Fraser University, Department of Physics, 
8888 University Drive, Burnaby, British Columbia, Canada V5A 1S6}
\thanks{Permanent address.}
\affiliation{TRIUMF, 4004 Wesbrook Mall, Vancouver, BC, V6T 2A3, Canada}

\author{R. M. Woloshyn}
\affiliation{TRIUMF, 4004 Wesbrook Mall, Vancouver, BC, V6T 2A3, Canada}

\begin{abstract}
Perturbative expansions of several small Wilson loops are computed through 
next-to-next-to-leading order in unquenched lattice QCD, from Monte Carlo 
simulations at weak couplings. This approach provides a much simpler
alternative to conventional diagrammatic perturbation theory, 
and is applied here for the first time to full QCD.
Two different sets of lattice actions are considered: one set
uses the unimproved plaquette gluon action together with
the unimproved staggered-quark action; the other set uses
the one-loop-improved Symanzik gauge-field action together with
the so-called ``asqtad'' improved-staggered quark action.
Simulations are also done with different numbers of dynamical fermions.
An extensive study of the systematic uncertainties is presented,
which demonstrates that the small third-order perturbative component 
of the observables can be reliably extracted from
simulation data. We also investigate the use of the rational 
hybrid Monte Carlo algorithm
for unquenched simulations with unimproved-staggered fermions.
Our results are in excellent agreement with diagrammatic perturbation 
theory, and provide an important cross-check of the perturbation
theory input to a recent determination of the strong coupling
$\alpha_{\overline{\rm MS}}(M_Z)$ by the HPQCD collaboration.
\end{abstract}

\maketitle

\section{\label{S:Intro}Introduction}

A key ingredient in many high-precision applications of lattice QCD is
the use of perturbation theory, in order to match lattice 
discretizations of actions and observables to their counterparts in 
continuum QCD. An important example is the determination of the strong 
coupling $\alpha_{\overline{\rm MS}}(M_Z)$,
where perturbative expansions of short-distance quantities,
such as small Wilson loops, can be used to extract the value of the 
coupling from simulation measurements \cite{NRQCDalphas,HPQCDalphas}. 
Lattice perturbation theory by Feynman diagram analysis is extremely 
difficult however, because the lattice regulator results in Feynman 
rules that are exceedingly complicated; moreover, there is a 
proliferation of diagrams that are not present in the continuum.
A further challenge is that these perturbative-matching calculations 
must generally be carried out at next-to-next-to-leading order 
(``NNLO,'' which is generally equivalent to two-loop Feynman 
diagrams), if one is to obtain results of a few-percent 
precision \cite{HDTReview}.

Although many parts of diagrammatic lattice perturbation theory have 
been automated with the help of computer codes 
\cite{LuscherVertex,HDTReview,QThesis}, 
higher-order calculations remain very challenging, and 
very few NNLO lattice calculations have been done (see, e.g.,
Refs.~\cite{UnimprovedPT,HPQCDalphas,HQalphaPT,HPQCDmass}). 
The algebraic burden is particularly heavy for the highly-improved 
actions that are now commonly used in numerical 
simulations. A particularly important case is the tree-level
$O(a^2)$-improved staggered-quark action (where $a$ is the lattice spacing) 
\cite{LepageStaggered}, together with the $O(a^2)$-accurate and one-loop
Symanzik-improved gluon action \cite{Symanzik,LuscherVertex}, both of 
which are tadpole improved (and which are hereafter collectively referred 
to as the ``asqtad'' actions). The ``asqtad'' actions are
currently being used by the MILC collaboration to generate unquenched 
ensembles with three-flavors of sea quarks \cite{MILCsims}, which
have been used by several groups for a wide variety of physics 
applications, including quantities requiring higher-order perturbation 
theory \cite{HPQCDalphas,HQalphaPT,HPQCDmass}.

Given the central role of perturbative matching in lattice QCD, 
alternatives to diagrammatic perturbation theory are desirable.
One approach \cite{Dimm,QuenchedHighBeta} is to use Monte Carlo 
simulations at weak couplings, where the theory enters the 
perturbative phase (at finite volume).
Simulation measurements of a particular observable are done at several 
values of the coupling, and the resulting data are fit 
to an expansion in the coupling; the fit yields numerical values for
the perturbative coefficients, without Feynman diagrams, and
with little or no analytic input.  

This Monte Carlo approach has been successful in computing the perturbation 
series of a number of quantities in pure gauge theories 
\cite{Dimm,QuenchedHighBeta,Juge,Hart}. In particular, perturbative 
coefficients for a number of small Wilson loops for the plaquette 
gluon action were computed to NNLO using this method in 
Ref.~\cite{QuenchedHighBeta}, and the results were subsequently 
reproduced by diagrammatic perturbation theory 
(see Ref.~\cite{HDTReview}). 

In this paper we make the first applications of this Monte Carlo method 
to lattice actions with dynamical fermions \cite{KitThesis}.
The perturbation series of a number of small Wilson loops are computed
through NNLO, for two different lattice QCD actions with dynamical 
staggered quarks: the Wilson plaquette gluon action with the 
unimproved-staggered quark action, and the ``asqtad'' actions. 
The latter calculation
is of particular relevance because it provides a consistency check
of the NNLO Wilson loop expansions used in the determination of 
$\alpha_{\overline{\rm MS}}(M_Z)$ by the HPQCD collaboration,
in Ref.~\cite{HPQCDalphas}.

We note that another simulation approach to lattice perturbation theory 
has also been developed, based on an explicit perturbative expansion 
of the simulation equations themselves \cite{RenzoQuenched}, and has 
recently been applied to (unimproved) lattice actions with 
dynamical fermions \cite{RenzoDynamical}.

The Monte Carlo method considered here has the advantage that it
can be done using conventional simulation codes, although it is
very advantageous to adopt twisted boundary conditions (TBC) 
\cite{tHooft,Parisi,LuscherVertex,Wohlert,Arroyo} in order to eliminate 
zero modes, and to suppress nonperturbative finite-volume effects due to 
transitions between $Z(3)$ phases \cite{QuenchedHighBeta}; 
fortunately, TBC can be implemented in existing simulation codes 
with relatively little effort.

This simulation approach to unquenched perturbation theory
is also very efficient, since the simulations can be done on 
very small lattices ($8^4$ volumes are used here), thanks to the use
of TBC. The simulations are also done here for different numbers $n_f$ of 
dynamical fermions, which explicitly demonstrates that the data are 
sensitive to the $n_f$-dependence of the perturbative coefficients.

Although small Wilson loops are the simplest observables for 
perturbative analysis by Monte Carlo simulations, because they
are such ultraviolet quantities (as well as being gauge invariant), 
the studies presented here 
nonetheless provide important benchmarks for future applications to
other quantities; for instance, similar methods were used to
analyze the static-quark propagator in pure-gauge backgrounds in
Ref.~\cite{QuenchedHighBeta}, and the Fermilab heavy-quark propagator
was analyzed in Ref.~\cite{Juge}.

While the expansion coefficients can be extracted with far less effort 
from Monte Carlo simulations, care must be taken to control all of 
the systematic uncertainties, in order to
reliably extract the small part of the signal corresponding to the 
higher-order terms in the perturbative expansion. It is also crucial to 
use an expansion in a renormalized coupling, rather than in the bare 
lattice coupling $\alpha_{\rm lat}=g^2/(4\pi)$, for which perturbation 
theory is very poorly convergent \cite{LepageMackenzie}. We use an expansion 
in the coupling $\alpha_V(q^*)$ defined by the static potential, along with 
an estimate of the optimal scale $q^*$ for a given quantity 
\cite{Brodsky,LepageMackenzie}. Although one can in principle define the 
renormalized coupling $\alpha_V$ at a given bare lattice coupling 
$\alpha_{\rm lat}$ from simulation measurements of the static potential, 
we rely here instead on existing NNLO determinations of the relationship 
from diagrammatic perturbation theory \cite{HQalphaPT}.

We did simulations at couplings $\alpha_V \alt 0.1$, hence 
statistical and systematic errors must be much less than 
$\rm{max}(\alpha_V^3) \sim 10^{-3}$, for determination of the third-order 
coefficients. The ensemble sizes were chosen in order to reduce the
statistical errors to the desired level.
There are four major sources of systematic 
error in the simulation algorithms and in the data analysis: 
i) transitions between the $Z(3)$ center phases of the gauge field action; 
ii) finite step-size errors in the simulation equations; 
iii) other algorithmic systematics, such as the precision of the
matrix inversion; and iv) fitting and truncation errors in
the perturbative expansion.

In this study simulations were done with TBC to 
suppress transition between the center phases, as indicated above.
To eliminate the finite step-size errors we pursued one of 
two strategies: in the case of the unimproved
actions, we used the rational hybrid Monte Carlo algorithm
(RHMC)~\cite{Horvath,Clark}; for the ``asqtad'' actions
we used the $R$ algorithm \cite{Gottlieb}, and did simulations 
at several different molecular-dynamics step sizes, the results of 
which were extrapolated to zero step size. To extract the
perturbative coefficients the data were analyzed 
using constrained curve fitting techniques \cite{LepageFit},
which provide an elegant procedure for incorporating our
{\it a priori\/} expectation that the perturbative expansion is
well behaved, and which readily allow the maximum amount of 
information to be extracted from the data.

The rest of the paper is organized as follows. The actions and
simulation parameters are detailed in Sect.~\ref{S:Actions},
along with the perturbation theory input which we use to extract
the renormalized coupling $\alpha_V$ at a given bare lattice coupling,
from the simulation data. A detailed analysis of the
systematic uncertainties is presented in Sect.~\ref{S:Systematics}, 
including the use of TBC to control finite-volume effects. Results for 
the two actions are reported in Sect.~\ref{S:Results}, and are compared 
with NNLO diagrammatic perturbation theory. Conclusions and prospects for 
further work are presented in Sect.~\ref{S:Conclusions}.

\section{\label{S:Actions}Actions and simulation parameters}

\subsection{Perturbative expansions}

We have done unquenched simulations for two sets of actions: the
first set used the unimproved Wilson plaquette action with 
unimproved-staggered fermions, while the second set used 
$O(a^2)$-improved gluon and staggered-quark actions. We provide the 
simulation parameters and perturbation theory input for each set 
of actions in the two following subsections.

A basic input parameter for the simulations is the bare lattice coupling
$\alpha_{\rm lat} = g^2/(4\pi)$, related to the usual simulation parameter 
$\beta$ in the case of the Wilson plaquette action, for example, according to 
$\beta \equiv 6/g^2 = 6/(4\pi\alpha_{\rm lat})$. However the bare coupling 
is not suitable for perturbative expansions \cite{LepageMackenzie} and
we use instead a renormalized coupling $\alpha_V(q)$, defined by the 
static potential according to \cite{Brodsky,LepageMackenzie}
\begin{align}
   V(q) \equiv -\frac43 \frac{4\pi\alpha_V(q)}{q^2} .
\label{alphaV}
\end{align}
A (truncated) perturbative expansion for the logarithm of a
Wilson loop of size $R \times T$ is used, owing to the perturbative 
perimeter law, 
\begin{align}
   -\frac{1}{2(R+T)} \ln W_{RT} \approx
   \sum_{n=1}^N c_{n;RT} \, \alpha_{V}^{n}(q_{RT}^*) ,
   \label{lnWRT}
\end{align}
where $N$ is the order at which the series is truncated. We aim to
measure the coefficients through third-order, hence fits must be
done with $N \ge 3$. The characteristic scale $q_{RT}^*$ for each 
observable is determined according to the BLM method (which estimates 
the typical momentum carried by a gluon in leading-order diagrams)
\cite{Brodsky,LepageMackenzie}. 

The connection between the simulation input $\alpha_{\rm lat}$ and 
the $\alpha_V$ coupling is therefore required. Although it is possible,
in principle, to extract this relation by directly measuring the 
static-quark potential in Monte Carlo simulations, the connection has 
already been computed in diagrammatic perturbation theory through 
NNLO for a variety of actions, including those we consider here, in 
Ref.~\cite{HQalphaPT}.
We use the results of Ref.~\cite{HQalphaPT} to provide the
three leading orders of the expansion for the $1\times 1$ plaquette,
as well as the scales $q_{RT}^*$ for all the Wilson loops (the
later only requires a relatively straightforward one-loop calculation).

The diagrammatic input provided by the expansion of the
$1\times1$ loop simplifies our analysis, and still allows for 
highly-nontrivial applications of the Monte Carlo method;
here we demonstrate the utility of the method by computing the NNLO
perturbative expansions of several larger Wilson loops, and thereby also 
provide a valuable consistency check of the NNLO diagrammatic calculations 
which were used to obtain $\alpha_{\overline{\rm MS}}(M_Z)$ in 
Refs.~\cite{HPQCDalphas,HQalphaPT}.

Given this input from diagrammatic perturbation theory, the Monte Carlo method 
proceeds as follows. Simulations are done at several small values of the bare 
coupling $\alpha_{\rm lat}$ (at which the lattice theory is in the perturbative 
phase at a given finite volume). For each bare coupling one measures the 
average plaquette $\langle W_{11}\rangle_{\mathrm{MC}}$, whose perturbative
expansion is assumed, and the quantities of interest, whose
perturbative expansions are to be determined; in our case these 
latter quantities are larger Wilson loops, 
$\langle W_{RT}\rangle_{\mathrm{MC}}$. 
The numerical value of the renormalized coupling $\alpha_{V}$ at the 
scale $q_{11}^*$ is obtained by substituting the measured value
$\langle W_{11}\rangle_{\mathrm{MC}}$ into its known third-order 
expansion. Once the numerical value of $\alpha_{V}(q_{11}^{*})$ has 
been determined, the couplings at the other relevant scales 
$q_{RT}^*$ can be computed using the universal second-order 
beta function, plus the known third-order correction in the 
$V$ scheme \cite{Schroder}, according to 
\begin{widetext}
\begin{align}
   \alpha_V(q) = \frac{4\pi}{\beta_{0}\ln(q^2/\Lambda_V^2)}
   \left[1- \frac{\beta_1}{\beta_0^2}
   \frac{\ln\left[\ln(q^2/\Lambda_V^2)\right]}{\ln(q^2/\Lambda_V^2)}
   + \frac{\beta_1^2}{\beta_0^4\ln^2(q^2/\Lambda_V^2)}
   \left( \left(\ln\left[\ln(q^2/\Lambda_V^2)\right] -\frac12 \right)^2  
   + \frac{\beta_{2V}\beta_0}{\beta_1^2} -\frac54 \right)\right], 
\label{LambdaV}
\end{align}
\end{widetext}
where $\alpha_V(q_{11}^{*})$ is traded for the intrinsic scale 
$\Lambda_V$ at the given bare coupling. The beta-function coefficients
are $\beta_0=11-\frac{2}{3}n_f$, $\beta_1=102-\frac{38}{3}n_f$, and
$\beta_{2V}=4224.18-746.006n_f+20.8719n_f^2$. Fits to the
expansion Eq.~(\ref{lnWRT}) then yield the perturbative coefficients for
the larger Wilson loops; details on the fitting procedure are
described in Sect.~\ref{S:Fitting}.

\subsection{Unimproved actions}

\begin{table*}
\begin{center}
\begin{tabular}{|c|c|c|c|c|c|c|}
\hline
\multicolumn{7}{|c|}{Number of flavors: $n_{f}=1$} \\ 
\hline
$\beta$                  & $\alpha_{\rm lat}=6/4\pi\beta$  &
$\langle W_{11}\rangle$ & Measurements              &
Acc.\ Rate                & $\alpha_V(q_{11}^{*})$ &
$a\Lambda_V$
\\ \hline
11.0 & 0.04341 & 0.805569(41) & 849  & 92\% & 0.05574 & $2.1\times 10^{-5}$  \\
13.5 & 0.03537 & 0.843964(34) & 902  & 90\% & 0.04292 & $9.1\times 10^{-6}$  \\
16.0 & 0.02984 & 0.869561(26) & 992  & 87\% & 0.03496 & $4.0\times 10^{-7}$  \\
19.0 & 0.02513 & 0.890999(21) & 981  & 81\% & 0.02860 & $9.2\times 10^{-9}$  \\
24.0 & 0.01989 & 0.914413(16) & 1066 & 69\% & 0.02198 & $1.7\times 10^{-11}$ \\
32.0 & 0.01492 & 0.936259(11) & 1180 & 58\% & 0.01605 & $7.2\times 10^{-16}$ \\
47.0 & 0.01016 & 0.956886(7)  & 1261 & 53\% & 0.01067 & $4.2\times 10^{-24}$ \\
\hline
\end{tabular}
\end{center}
\begin{center}
\begin{tabular}{|c|c|c|c|c|c|c|}
\hline
\multicolumn{7}{|c|}{Number of flavors: $n_{f}=3$} \\ 
\hline
$\beta$                  & $\alpha_{\rm lat}=6/4\pi\beta$  &
$\langle W_{11}\rangle$ & Measurements              &
Acc.\ Rate                & $\alpha_V(q_{11}^{*})$ &
$a\Lambda_V$
\\ \hline
11.0 & 0.04341 & 0.807064(36) & 821  & 89\% & 0.05551 & $3.9\times 10^{-6}$  \\
13.5 & 0.03537 & 0.844847(26) & 641  & 86\% & 0.04283 & $1.1\times 10^{-6}$  \\
16.0 & 0.02984 & 0.870205(21) & 740  & 84\% & 0.03490 & $2.9\times 10^{-8}$  \\
19.0 & 0.02513 & 0.891387(18) & 738  & 77\% & 0.02859 & $3.8\times 10^{-10}$ \\
24.0 & 0.01989 & 0.914647(13) & 791  & 66\% & 0.02197 & $2.7\times 10^{-13}$ \\
32.0 & 0.01492 & 0.936402(10) & 840  & 56\% & 0.01605 & $2.5\times 10^{-18}$ \\
47.0 & 0.01016 & 0.956950(6)  & 902  & 52\% & 0.01066 & $8.5\times 10^{-28}$ \\
\hline
\end{tabular}
\caption{Simulation parameters for the unimproved plaquette gluon action
with the unimproved staggered-quark action. Two different sets of simulations 
were done for the indicated number of flavors; in both cases $8^4$ 
lattices with twisted boundary conditions were used, with bare-quark mass 
$m_0a=0.2$. The RHMC algorithm was used with step size $\Delta t=0.01$, with 
50 molecular-dynamics steps per trajectory; the table shows the
acceptance rate for the accept/reject step at the end of each trajectory.}
\label{UnimpParams}
\end{center}
\end{table*}

\begin{table}
\centering
\begin{tabular}{|c|c||c|c|}
\hline
Loop & $q_{RT}^{*}$ & Loop & $q_{RT}^{*}$ \\ \hline
$1\times 1$ & 3.40  & $2\times 2$ & 2.65    \\
$1\times 2$ & 3.07  & $2\times 3$ & 2.56    \\
$1\times 3$ & 3.01  & $3\times 3$ & 2.46    \\ \hline
\end{tabular}
\caption{The optimal momentum scales $q_{RT}^{*}$ for selected
$R\times T$ Wilson loops for the unimproved actions \cite{HQalphaPT}.}
\label{UnimpQstar}
\end{table}

The first set of actions we consider are the Wilson plaquette gluon action,
\begin{equation}
   S^{\rm unimp}_ {\rm glue} = \beta \sum_{x;\mu < \nu} (1-P_{\mu\nu}) ,
\label{SUnimpGlue}
\end{equation}
where $P_{\mu\nu}$ is the $1\times1$ plaquette and $\beta = 6/g^2$, 
together with the unimproved staggered-quark action
\begin{equation}
   S^{\rm unimp}_{\rm stagg} = \sum_x \bar\chi(x) \left( \sum_\mu
   \eta_\mu(x) \Delta_\mu + m_0 \right) \chi(x)  ,
\label{SUnimpNaive}
\end{equation}
where $\Delta_\mu$ is the standard lattice derivative,  
and $\eta_\mu(x) = (-1)^{x_1 + \ldots + x_{\mu-1}}$ is 
the usual staggered-quark phase.

Two sets of simulations were done for the unimproved actions; the 
simulation parameters are summarized in Table~\ref{UnimpParams}.
One set was done for a single dynamical flavor, $n_f = 1$, and another 
was done for three degenerate dynamical flavors, $n_f=3$. In each case, 
simulations were done for seven values of the bare coupling, with 
bare-quark mass $m_0a=0.2$, on $8^4$ lattices with twisted boundary 
conditions (the boundary conditions are discussed in Sect.~\ref{S:TBC}).
The configurations for these actions were generated using the RHMC algorithm,
described in more detail in Sect.~\ref{S:RHMC}; we used a time-step 
$\Delta t=0.01$ with 50 molecular-dynamics steps between accept/reject tests. 
Based on an autocorrelation analysis, described in 
Sect.~\ref{S:OtherSystematics}, 
40 trajectories were skipped between measurements.

For the unimproved actions the expansion coefficients of the average 
plaquette to NNLO are given by \cite{HQalphaPT,PlaqPTNote}
\begin{align}
   c_{1;11} & = 1.04720(0) ,
\nonumber \\
   c_{2;11} & = -1.2467(2) - 0.06981(5) n_f ,
\nonumber \\
   c_{3;11} & = -1.778(7) + 0.464(27)n_f + 0.00485(0) n_f^2 ,
\label{Unimpc11}
\end{align}
where the uncertainties in the coefficients are statistical errors
which arise from numerical integration of the multi-loop Feynman
diagrams using a Monte Carlo technique (the error of ``0'' in
$c_{1;11}$ indicates that the integration error is 
in the sixth digit in that case). The relevant scales for
the Wilson loops are given in Table~\ref{UnimpQstar}.

\subsection{Improved actions}

\begin{table*}
\begin{center}
\begin{tabular}{|c|c|c|c|c|c|c|}
\hline
$\beta$   & Input $u_0$   & Measured $u_0$   &  $\alpha_{\rm lat}$
          & $\alpha_V(q^*_{11})$     &  $a\Lambda_V$   & Measurements \\
\hline
9.5  & 0.91690 & 0.916922(98) & 0.08377	& 0.12709 & $5.8 \times 10^{-2}$ &
461,457,836,1335 \\
11.0 & 0.93166 & 0.931687(73) & 0.07234 & 0.10113 & $2.0 \times 10^{-2}$ &
281,633,1090,1038 \\
13.5 & 0.94704 & 0.946986(55) & 0.05895 & 0.07604 & $3.2 \times 10^{-3}$ &
290,635,1122,1078 \\
16.0 & 0.95661 & 0.956614(53) & 0.04974 & 0.06106 & $5.0 \times 10^{-4}$ &
296,634,1145,1100 \\
19.0 & 0.96433 & 0.964311(38) & 0.04188 & 0.04951 & $5.5 \times 10^{-5}$ &
298,643,1123,1172 \\
24.0 & 0.97243 & 0.972453(29) & 0.03316 & 0.03765 & $1.3 \times 10^{-6}$ &
308,645,1191,1113 \\
32.0 & 0.97978 & 0.979785(23) & 0.02487 & 0.02727 & $3.3 \times 10^{-9}$ &
315,652,1205,1204 \\
47.0 & 0.98652 & 0.986526(15) & 0.01693 & 0.01797 & $3.8 \times 10^{-14}$ &
215,709,1227,963  \\
80.0 & 0.99220 & 0.992208(13) & 0.00995 & 0.01029 & $5.3 \times 10^{-25}$ &
472,489,974,1527 \\
\hline
\end{tabular}
\caption{Simulation parameters for the ``asqtad'' actions on $8^4$ 
lattices with twisted boundary conditions.
The number of flavors is $n_f=1$, with bare-quark mass $m_0a=0.1$.
Simulations at each bare lattice coupling were done at four values of 
the $R$-algorithm step-size, $\Delta t = 0.005, 0.01, 0.02$, and 0.03. 
The measured $u_0$ and $\alpha_V(q^*_{11})$ are only shown for the 
ensembles with $\Delta t = 0.005$. The number of measurements at each 
of the four step sizes is given.}
\label{ImpParams}
\end{center}
\end{table*}

\begin{table}
\centering
\begin{tabular}{|c|c||c|c|}
\hline
Loop & $q_{RT}^{*}$ & Loop & $q_{RT}^{*}$ \\ \hline
$1\times 1$ & 3.33   & $2\times 2$ & 2.58   \\
$1\times 2$ & 3.00   & $2\times 3$ & 2.48   \\
$1\times 3$ & 2.93   & $3\times 3$ & 2.37   \\ \hline
\end{tabular}
\caption{The optimal momentum scales $q_{RT}^{*}$ for selected
$R\times T$ Wilson loops for the ``asqtad'' actions \cite{HQalphaPT}.}
\label{ImpQstar}
\end{table}

The one-loop Symanzik-improved gluon action 
\cite{Symanzik,LuscherVertex} we use follows 
Refs.~\cite{Alford,MILCsims} for tadpole improvement 
\begin{align}
S^{\rm imp}_{\rm glue} & = \beta_{\rm pl} \sum_{x;\mu < \nu} (1-P_{\mu\nu})
   + \beta_{\rm rt} \sum_{x;\mu \ne \nu} (1-R_{\mu\nu}) 
   \nonumber \\
   & + \beta_{\rm pg} \sum_{x;\mu < \nu < \sigma}
   (1-C_{\hat\mu,\pm\hat\nu,\pm\hat\sigma}),
\label{SImpGlue}
\end{align}
where $R_{\mu\nu}$ is the $1\times 2$ rectangle, 
$C_{\mu\nu\sigma}$ is the $1\times1\times1$ ``corner cube''
(see Ref.~\cite{MILCsims}), and where the couplings are given by
\begin{align}
   & \beta_{\rm pl} = \frac{10}{g^{2}} , \quad
   \beta_{\rm rt} = - \frac{\beta_{\rm pl}}{20u_0^2} (1+0.4805\alpha_s) ,
\nonumber \\
   & \ \quad \beta_{\rm pg} = -\frac{\beta_{\rm pl}}{u_0^2} 0.03325\alpha_s ,
\end{align}
with $\alpha_s$ here defined by the (first-order accurate) expression
\begin{equation}
   \alpha_s \equiv -4 \ln(u_0) / 3.0684 .
\end{equation}
The one-loop couplings in $S^{\rm imp}_{\rm glue}$ correspond to the 
average-plaquette definition of the gluon mean field
\begin{equation}
   u_0 \equiv (W_{11})^{1/4} .
\end{equation}

The tree-level $O(a^2)$-accurate improved staggered-quark action we 
simulate was derived in Ref.~\cite{LepageStaggered}, and is also used in 
the three-flavor simulations by the MILC collaboration \cite{MILCsims}
\begin{align}
   S^{\rm imp}_{\rm stagg} = \sum_x \bar\chi(x) \left[ \sum_\mu
   \eta_\mu(x) \left( \Delta'_\mu - \frac{a^2}{6} \Delta^3_\mu \right) 
   + m_0 \right] \chi(x) ;
\label{SImpNaive}
\end{align}
see Ref.~\cite{LepageStaggered} for the definition of the ``smeared'' 
derivative operator $\Delta'_\mu$. Tadpole improvement of 
$S^{\rm imp}_{\rm stagg}$ is defined by replacing each link $U_\mu$ in the 
action by $U_\mu / u_0$, but only after adjacent pairs of 
identical links ($U^\dagger_\mu U_\mu = I$) are eliminated; 
the tadpole weights for the individual link paths in the quark 
action were also detailed by the MILC collaboration \cite{MILCsims}. 

Simulations were done for the ``asqtad'' actions only for a single
dynamical flavor, $n_f = 1$. Ensembles were generated at nine 
values of the bare coupling, on $8^4$ lattices and with bare-quark 
mass $m_0a=0.1$; simulation parameters are given in Table~\ref{ImpParams}.

The $R$ algorithm was used in these simulations, and ensembles
were generated at each bare coupling at four different 
molecular-dynamics step sizes, $\Delta t=0.005, 0.01, 0.02$, and 0.03,
with the number of steps per trajectory in the four cases given by 
100, 50, 25, and 15, respectively. The measured Wilson loop data are
extrapolated to $\Delta t=0$ at each bare coupling using constrained curve 
fitting (cf.\ Sect.~\ref{S:Fitting}).
The simulation value of the tadpole factor $u_0$ is determined 
self-consistently at each bare coupling by iteration during the 
thermalization process; the ``final'' input value was verified 
for consistency with the final measured value of $u_0$, as shown 
in Table~\ref{ImpParams}. Based on an autocorrelation analysis, 
described in Sect.~\ref{S:OtherSystematics}, we skipped 
20 trajectories between measurements.

For the ``asqtad'' actions the expansion coefficients of the average plaquette 
to NNLO are given by \cite{HQalphaPT,PlaqPTNote}
\begin{align}
   c_{1;11} & = 0.76710(0) ,
\nonumber \\
   c_{2;11} & = -0.5945(2) - 0.07391(2) n_f ,
\nonumber \\
   c_{3;11} & = -0.589(38) + 0.600(2)n_f + 0.00774(0) n_f^2 ,
\label{Impc11}
\end{align}
The relevant scales for the Wilson loops are given in Table~\ref{ImpQstar}.

\section{\label{S:Systematics}Systematic Effects}

The various systematic effects which must be controlled in order to
extract higher-order expansion coefficients were enumerated in the 
Introduction, and in the following four subsections we consider 
these in turn.

\subsection{\label{S:TBC}$Z(3)$ phases and twisted boundary conditions}

\begin{figure*}
\begin{center}
\begin{minipage}{2.5in}
\includegraphics[width=2.5in]{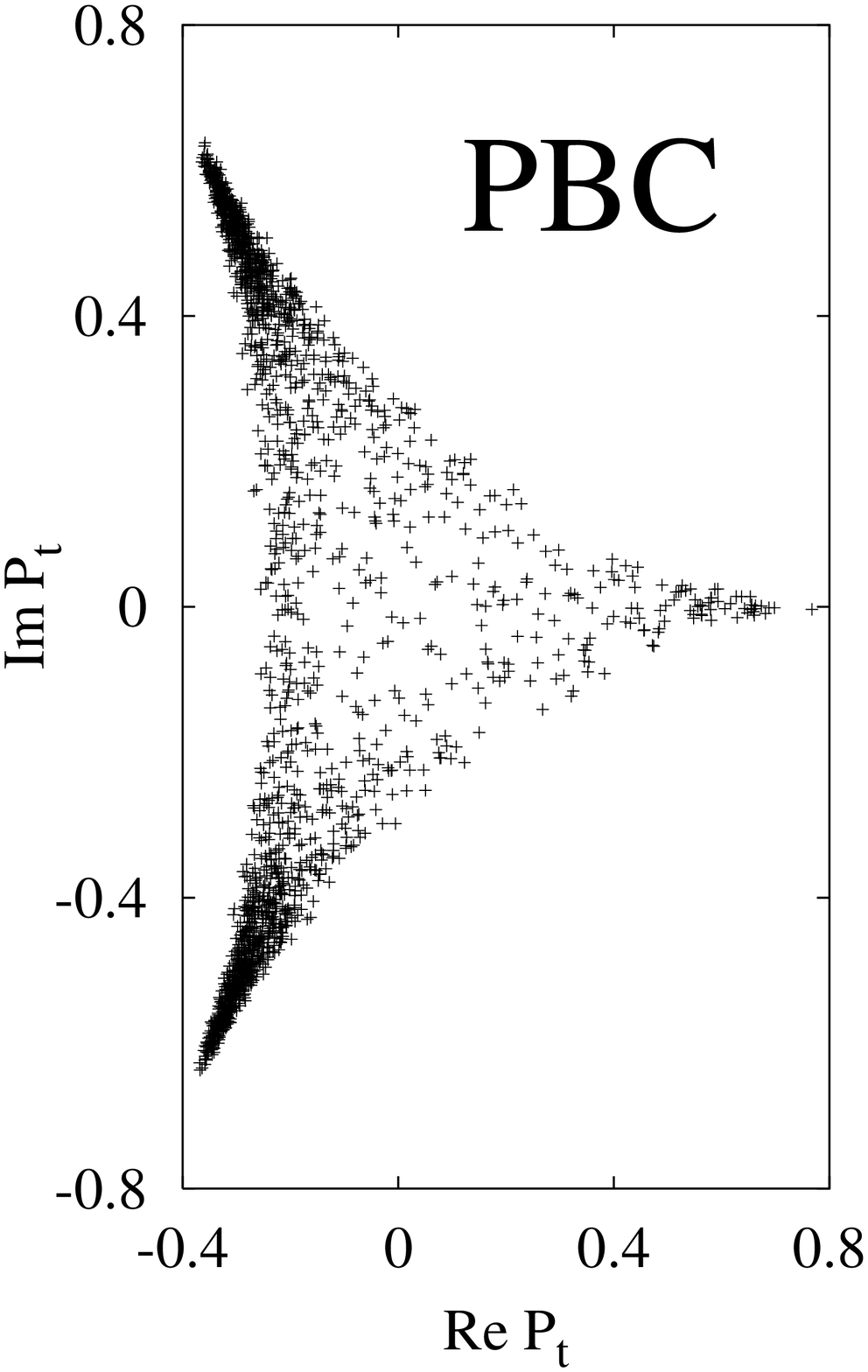}
\end{minipage}%
\begin{minipage}{2.5in}
\begin{minipage}{2.5in}
\includegraphics[width=2.5in]{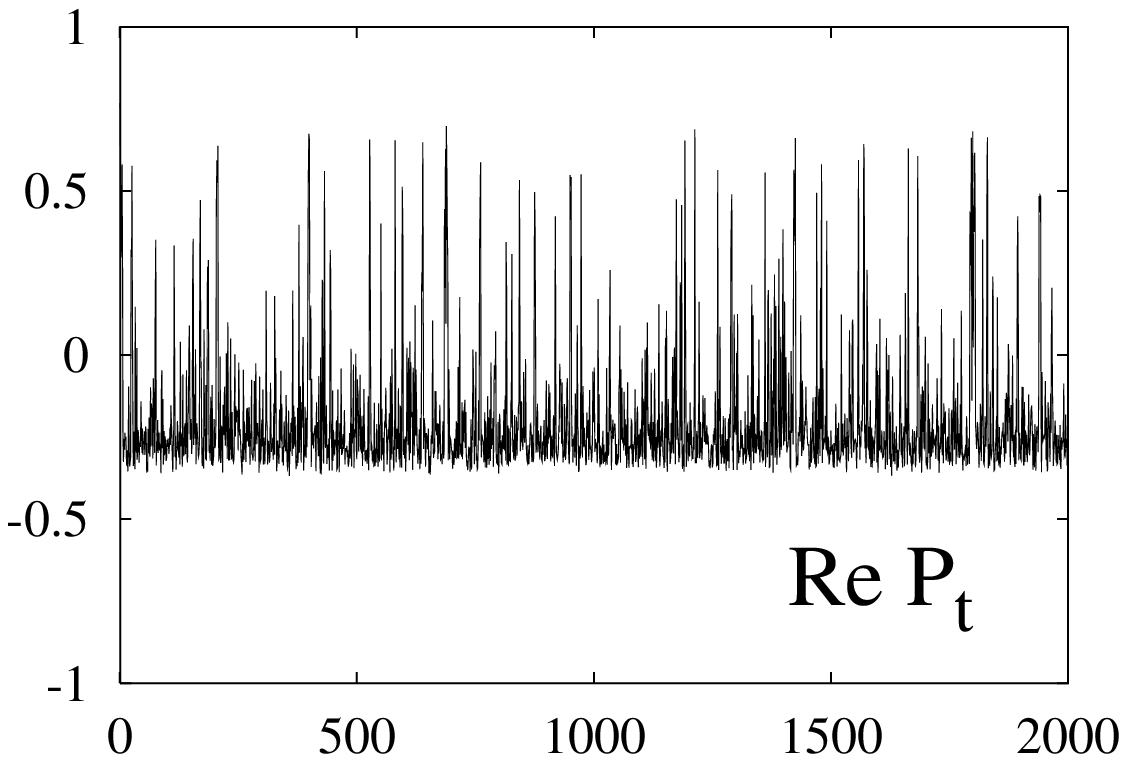}
\end{minipage}\\
\begin{minipage}{2.5in}
\includegraphics[width=2.5in]{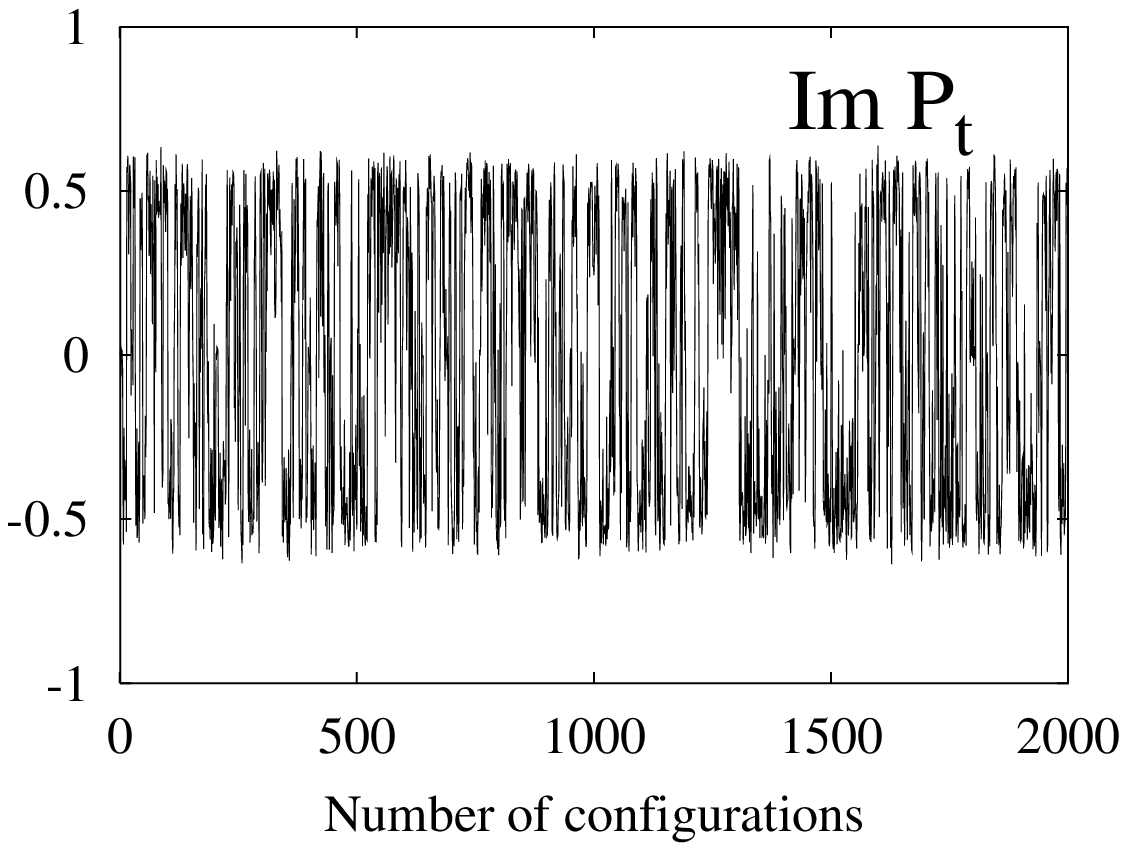}
\end{minipage}
\end{minipage}
\caption{Scatter plot and run-time history of the ``temporal''
Polyakov loop $P_t$ on a $4^4$ lattice at $\beta=16$, using
periodic boundary conditions. Results are shown for the
unimproved actions using the RHMC algorithm, with $n_f=1$, 
and bare-quark mass $m_0a=0.2$; 100 trajectories were skipped between 
measurements. The simulations were started with all links set to
the identity, $U_\mu(x) = I$, and the history is shown starting 
from this initial configuration.}
\label{fig:PBC}
\end{center}
\end{figure*}

\begin{figure*}
\begin{center}
\begin{minipage}{2.5in}
\includegraphics[width=2.5in]{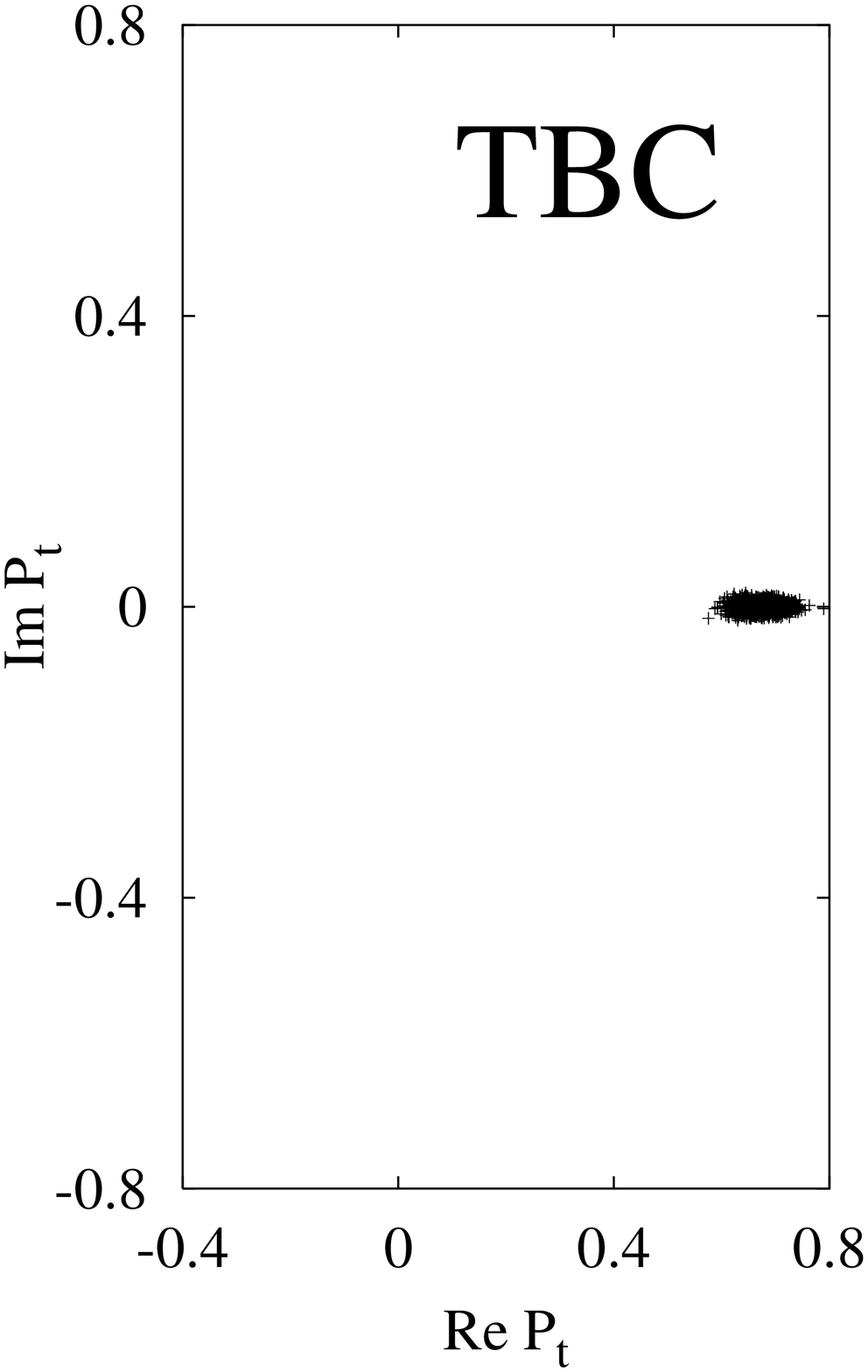}
\end{minipage}%
\begin{minipage}{2.5in}
\begin{minipage}{2.5in}
\includegraphics[width=2.5in]{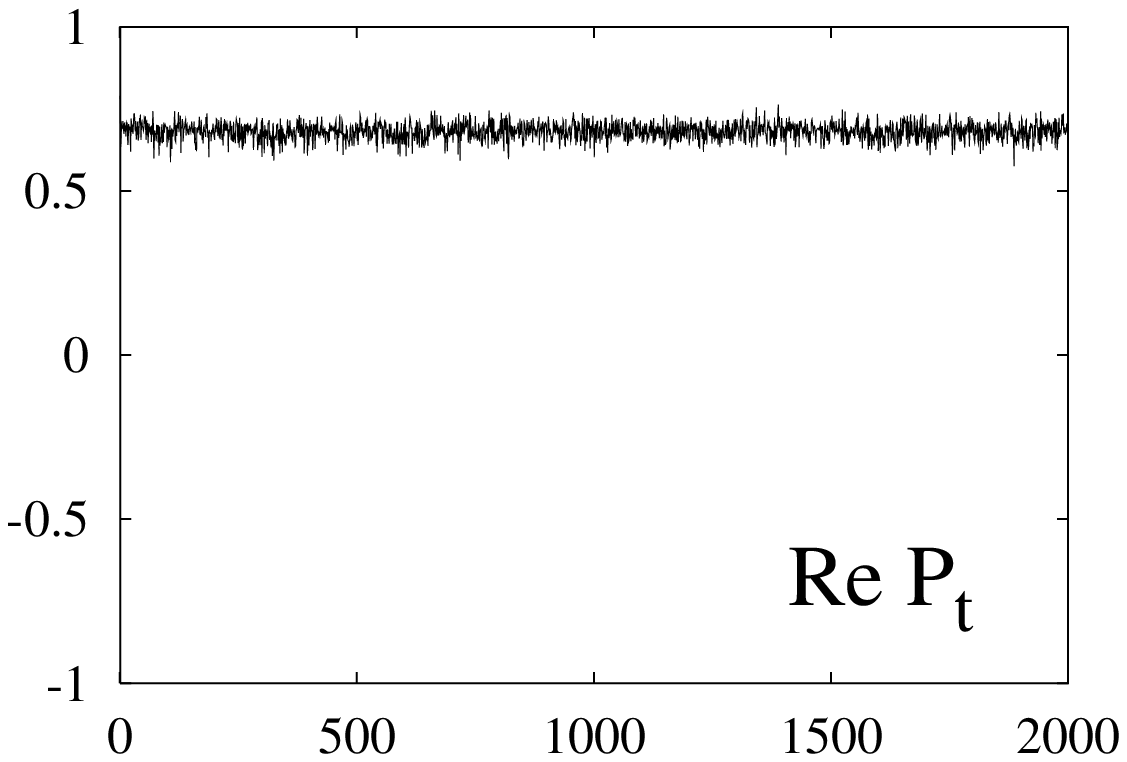}
\end{minipage}\\
\begin{minipage}{2.5in}
\includegraphics[width=2.5in]{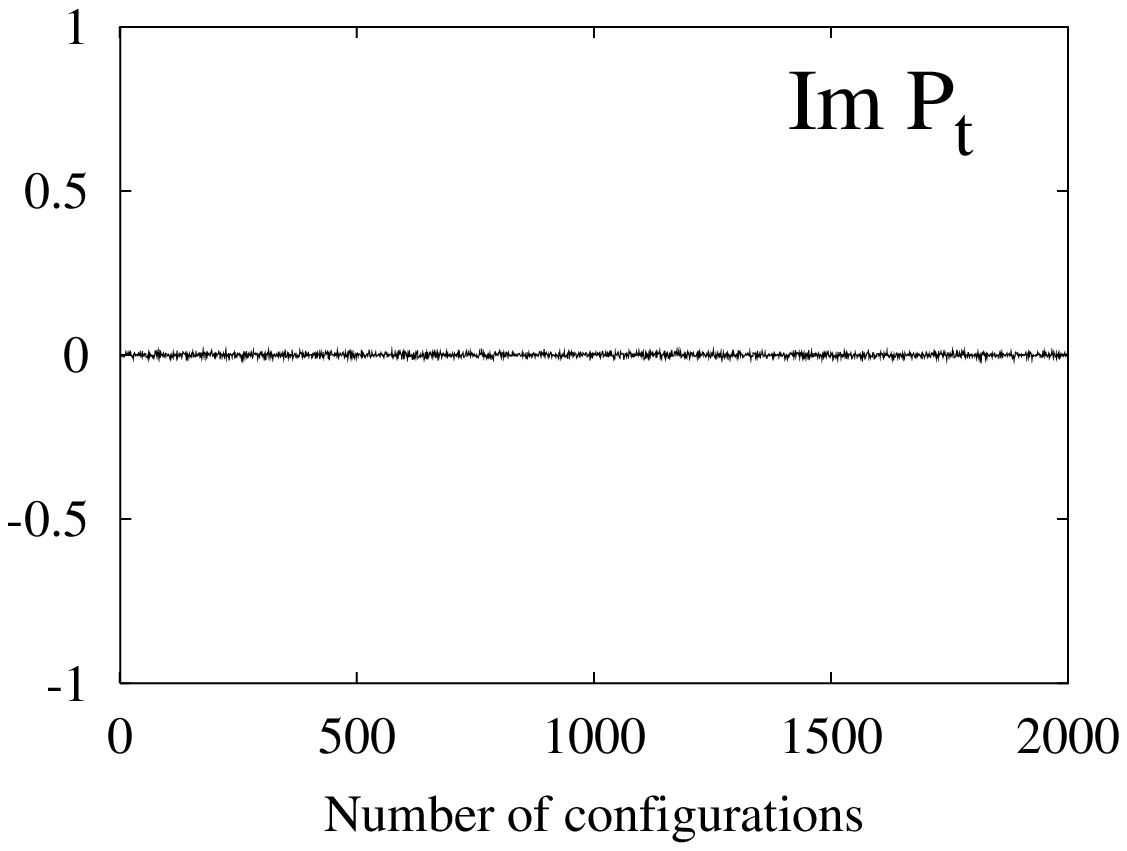}
\end{minipage}
\end{minipage}
\caption{Unimproved action scatter plot and run-time history,
with the same simulation parameters as in Fig.~\ref{fig:PBC}, but 
here using twisted boundary conditions (TBC) in the three
``spatial'' directions.}
\label{fig:TBC}
\end{center}
\end{figure*}

The SU(3)${}_{\rm color}$ gauge-field action is invariant under 
the transformation 
\begin{equation}
   U_\mu(x) \to z U_\mu(x),
   \quad \forall x \ni x \cdot \hat\mu = \mbox{constant} ,
\label{Z3}
\end{equation}
where $z \in [1,e^{i2\pi/3},e^{i4\pi/3}]$ is an element of 
the center-subgroup $Z(3)$. 
Although all closed Wilson loops are invariant under this transformation,
the Polyakov line in the $\mu$ direction is sensitive to the phase;
moreover, the perturbative expansion of Wilson loops can be spoiled
by the formation of domains of different center phases \cite{Dimm}.
While the fermion action breaks the center symmetry, the action ``cost'' 
for reaching different phases, at the lattice volume considered here, 
is too small to prevent frequent transitions between phases, at
least when periodic boundary conditions (PBC) are used.
This can be seen in Fig.~\ref{fig:PBC}, where a scatter
plot and run-time history of the ``temporal'' Polyakov line are shown
for the unimproved actions with PBC, on a $4^4$ lattice
at $\beta=16$, well into the deconfined phase of the theory.  

Although the scatter plot shows that the action with dynamical
fermions has a preference for configurations near the non-trivial 
phases $z=e^{i2\pi/3}$ and $z=e^{i4\pi/3}$ (despite the fact
that the simulation was started with all links set to the identity),
owing to the breaking of the $Z(3)$ symmetry, 
transitions between the various phases occur frequently, 
which would prevent the extraction of perturbative 
expansions, except on very large lattices \cite{QuenchedHighBeta}.

Fortunately, the $Z(3)$ transitions can be largely eliminated by using 
twisted boundary conditions (TBC) \cite{QuenchedHighBeta}, which can be 
easily implemented using existing simulation codes. In the case of the
gauge fields, there is no change to the link variables inside the
lattice, and link variables across the twisted lattice boundaries are
computed according to
\begin{equation}
   U_\mu(x+L\hat\nu) = \Omega_\nu U_\mu(x) \Omega_\nu^\dagger,
   \label{TBCU}
\end{equation}
where $L$ is the lattice length, and the $\Omega_\nu$ are a set of 
constant ``twist'' matrices which obey the algebra \cite{LuscherVertex}
\begin{equation}
   \Omega_\mu\Omega_\nu = z \, \Omega_\nu\Omega_\mu, \quad
   \Omega_\nu^3 \propto I .
\label{Omega}
\end{equation}
The gauge action and observables are therefore periodic with TBC, 
but with period $3L$, and with the important additional benefit that 
zero modes are eliminated \cite{LuscherVertex}.

Twists must be applied across at least two boundaries, else the effect of 
the twist matrix can be removed by a field redefinition. We choose
to impose twists across all three ``spatial'' lattice boundaries, with
ordinary periodic boundary conditions across the ``temporal''
boundary \cite{QuenchedHighBeta}. The following explicit representation 
of the twist matrices was used in our simulations
\begin{align}
   \Omega_{x} & = \left[ 
   \begin{array}{ccc}
   0 & 1 & 0 \\
   0 & 0 & 1 \\
   1 & 0 & 0
   \end{array} \right], \quad
   \Omega_{y} = \left[ 
   \begin{array}{ccc}
   e^{-i2\pi/3} & 0 & 0 \\
   0 & 1 & 0 \\
   0 & 0 & e^{+i2\pi/3}
   \end{array} \right],
\nonumber \\
   & \Omega_{z} = \Omega_{x}^{2}\Omega_{y} = \left[ 
   \begin{array}{ccc}
   0 & 0 & e^{-i2\pi/3} \\
   1 & 0 & 0 \\
   0 & e^{+i2\pi/3} & 0
   \end{array} \right].
\end{align}

To apply twists to the unquenched theory the quark field fields $\chi(x)$ 
must become 3$\times$3 color matrices \cite{Parisi,Wohlert}; these additional 
fermionic degrees-of-freedom amount to a new set of three degenerate 
``flavors'' (Parisi introduced the term ``smells'' \cite{Parisi}, to 
distinguish these copies from physical quark flavors).
To compute the fermion fields across the twisted lattice boundaries one
then imposes the boundary conditions 
\begin{equation}
   \chi(x+L\nu) = e^{i\pi/3} \Omega_\nu \chi(x) \Omega_\nu^\dagger ,
\label{fermionTBC}
\end{equation}
where the phase $e^{i\pi/3}$ makes the quark field anti-periodic 
(on intervals $3L$), thereby eliminating its zero modes as well.

As a result of the three-fold increase in the number of components of
the quark field, unquenched simulations using TBC will be roughly 
three times as expensive as simulations using PBC for the same action 
(although this cost is far offset by the reduction 
in finite volume effects). To compensate for the extra fermion copies 
we include an additional factor of $1/3$ in the fermion force term 
(on top of the $1/4$ to reduce the four staggered ``tastes'' to 
a single effective quark flavor); we therefore continue to use 
$n_f$ to denote the total number of quark ``flavors.''

Figure~\ref{fig:TBC} shows a scatter plot and run-time
history of the temporal Polyakov for simulations done with
TBC, using the same simulation parameters that were used with
PBC to generate Fig.~\ref{fig:PBC}. The suppression of
$Z(3)$ transitions when TBC are used is striking.
In fact, we have generated about one million configurations with TBC
(at a larger volume $V=8^4$), and not a single transition away from
the ``trivial'' $Z(3)$ phase has been observed.

\subsection{\label{S:RHMC}Simulation algorithms}

\begin{figure*}
\begin{center}
\hspace{-0.75in}
\begin{minipage}{2.25in}
\begin{minipage}{2.25in}
\includegraphics[width=2.25in,angle=-90]{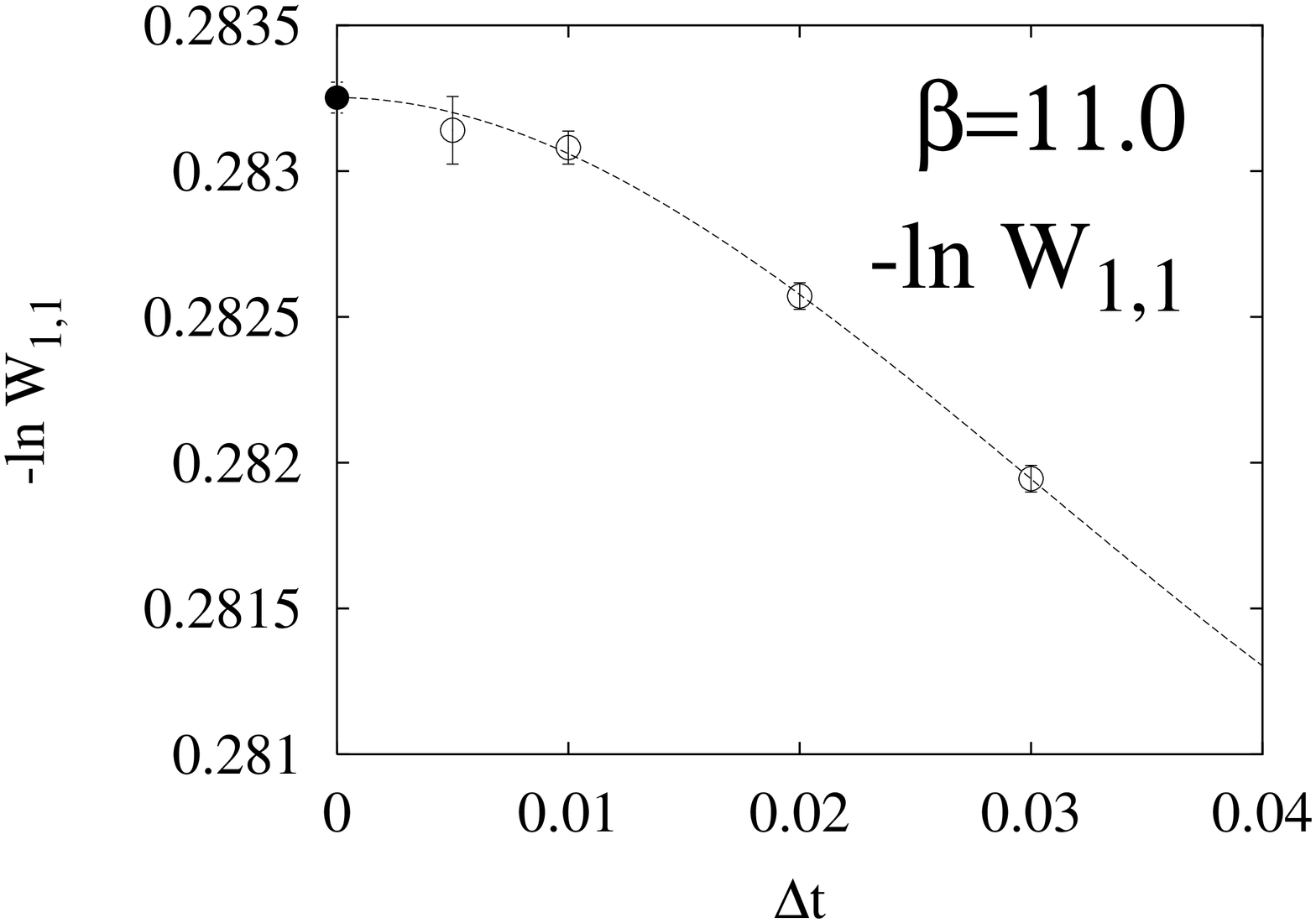}
\end{minipage} \\
\begin{minipage}{2.25in}
\includegraphics[width=2.25in,angle=-90]{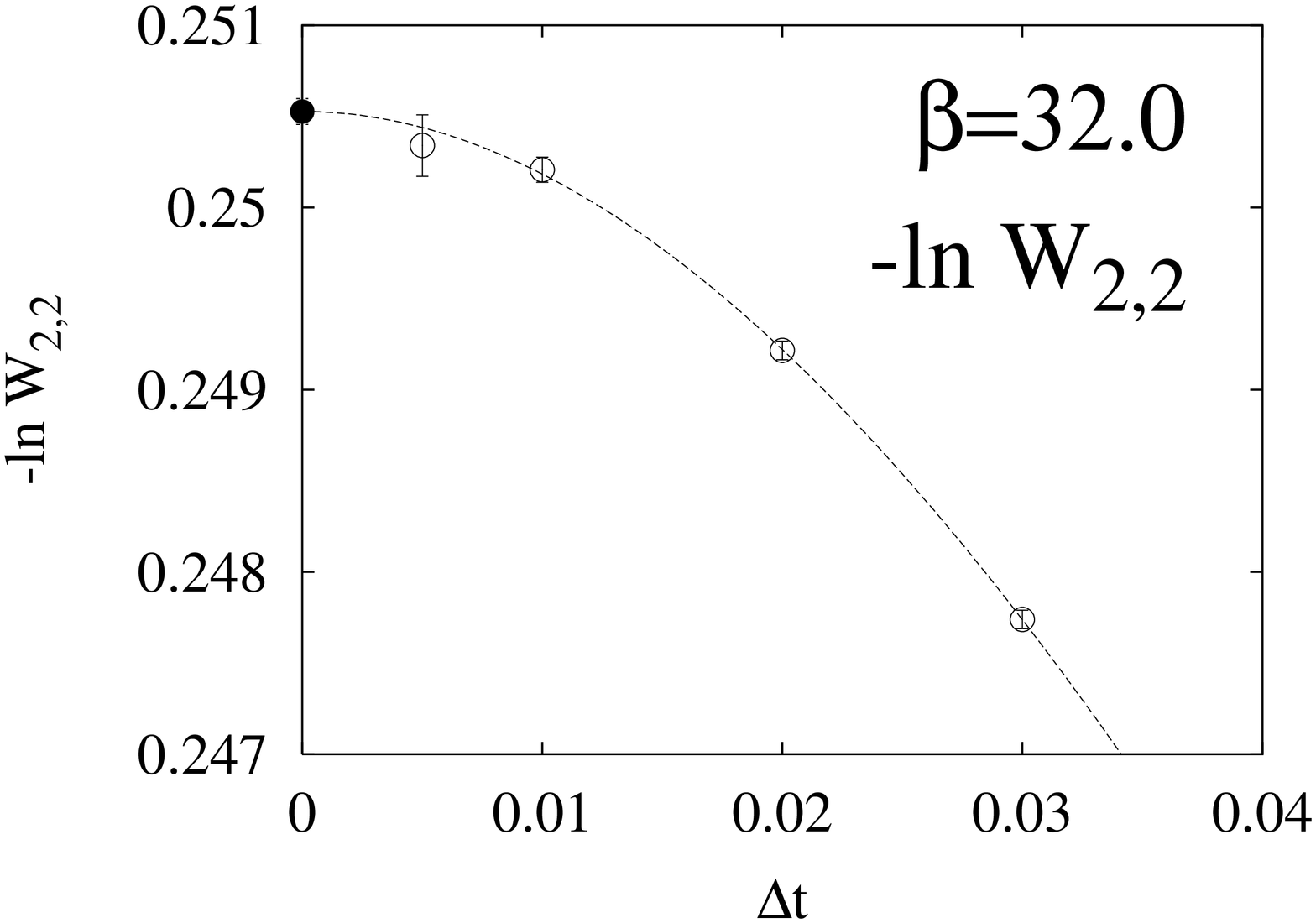}
\end{minipage}
\end{minipage}
\hspace{0.75in}
\begin{minipage}{2.25in}
\begin{minipage}{2.25in}
\includegraphics[width=2.25in,angle=-90]{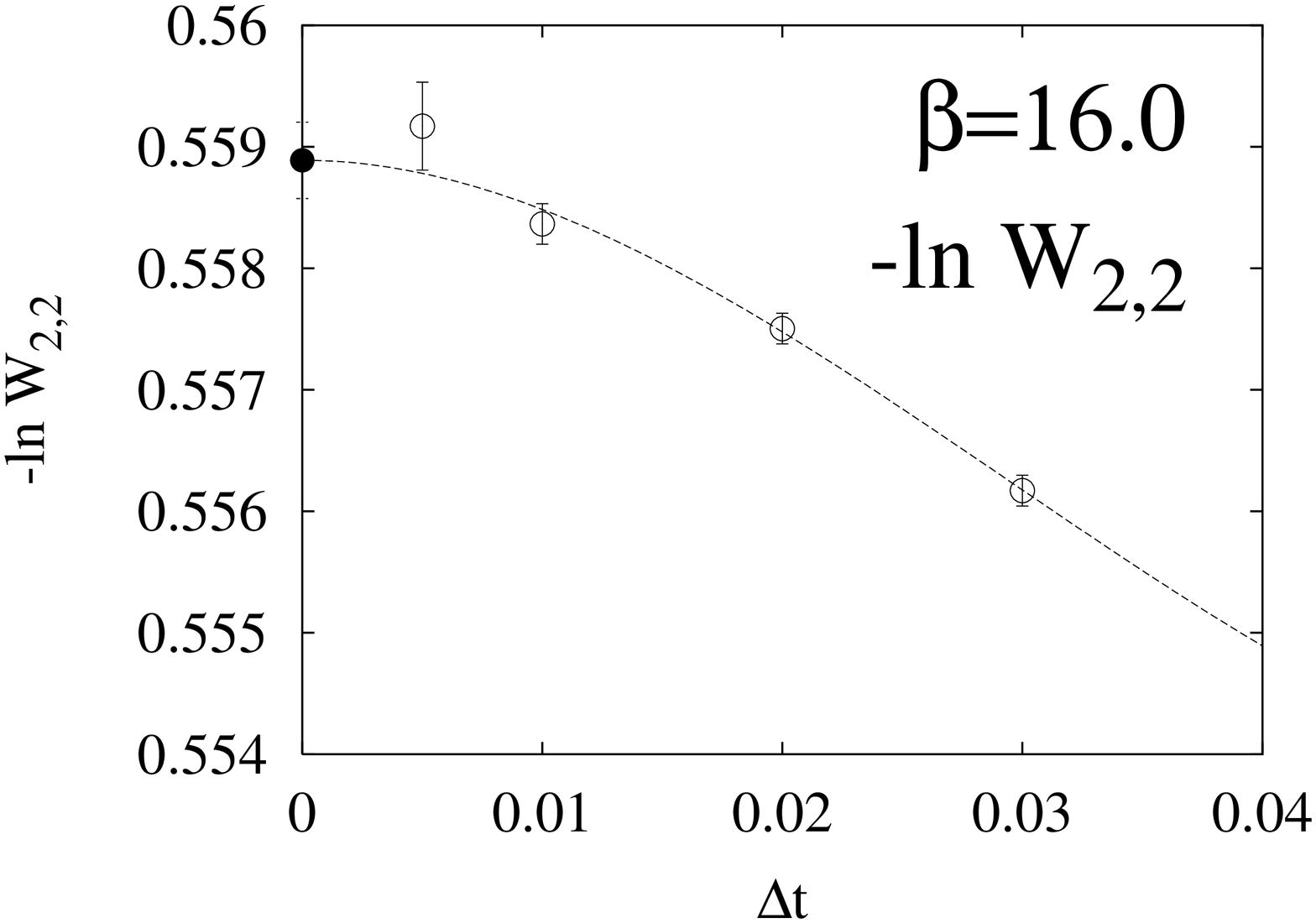}
\end{minipage} \\
\begin{minipage}{2.25in}
\includegraphics[width=2.25in,angle=-90]{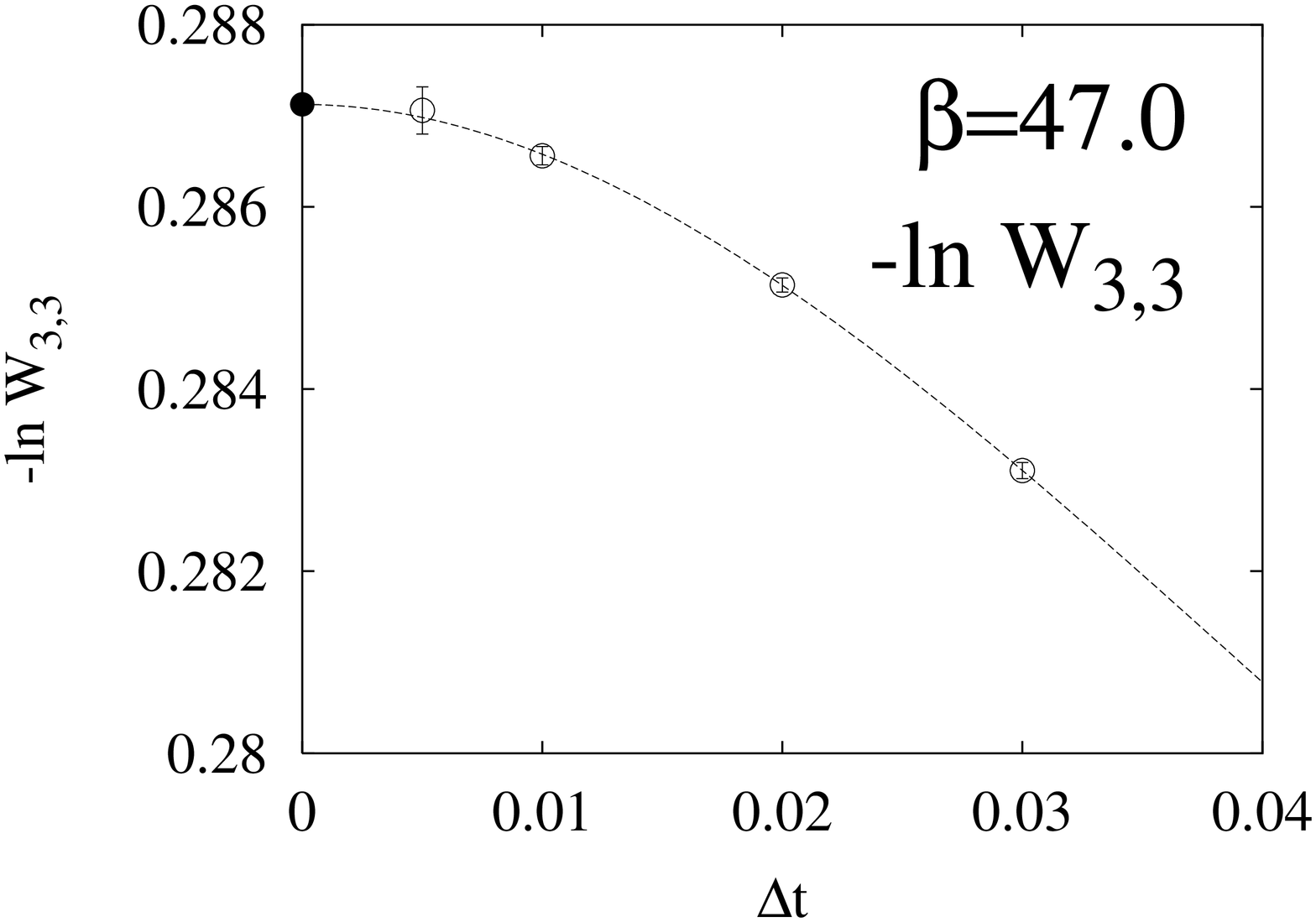}
\end{minipage}
\end{minipage}
\caption{Representative step-size extrapolations of the 
$R$-algorithm simulation results for the ``asqtad'' actions.}
\label{fig:Deltat}
\end{center}
\end{figure*}

Dynamical simulations with staggered quarks have generally 
been done with the $R$ algorithm. The major disadvantage of this
algorithm is that measured quantities have leading $O(\Delta t^2)$
errors \cite{Gottlieb}, where $\Delta t$ is the step-size in the
molecular-dynamics evolution equations. The step-size errors cannot be
corrected by a Monte Carlo accept/reject procedure, because
the fermion force is not computed explicitly in the $R$ algorithm, 
rather a noisy estimator is used. This leads to a large change in
the system energy during the molecular-dynamics updates,
making the acceptance rate small if one would impose a Monte Carlo 
accept/reject step at the end of the evolution. Therefore 
simulations are generally done at several values of $\Delta t$, 
and the results are extrapolated 
to zero step size. This is the method we use to simulate the 
``asqtad'' actions, for which ensembles were generated at four values of
$\Delta t$ at each bare lattice coupling (see Table~\ref{ImpParams}).
The Wilson loop data at a given bare coupling were fit to an expansion 
in $\Delta t$, including terms up to $O(\Delta t^6)$, excluding the linear 
term (constrained-curve fits were used, with priors for each term in the 
fit set to $0\pm 5$, see Sect.~\ref{S:Fitting}).
Some typical results for the step-size extrapolations
are shown in Fig.~\ref{fig:Deltat}.

Step-size errors can be eliminated using the rational hybrid Monte Carlo 
(RHMC) algorithm to handle the fractional powers of the staggered matrix
\cite{Horvath,Clark}.
In the RHMC algorithm the fermion force for staggered quarks is 
computed explicitly by approximating the $(n_f/4)$-th root (or the 
$(n_f/12)$-th root in the case of TBC) of the fermion matrix 
with a rational function (and retaining only even-site couplings in  
$M^{\dag}M$, as usual, to avoid a further redoubling of 
fermion degrees-of-freedom). 
This allows an explicit computation of the force term, hence 
the step-size errors can be corrected efficiently using a Monte Carlo 
accept/reject step. 

We tested the efficiency of the RHMC algorithm
by applying rational function approximations of various degrees
[$N_{\rm rat}$,$N_{\rm rat}$], that is, the numerator and denominator 
are taken to be polynomials of degree $N_{\rm rat}$. This was done
to approximate both $x^{-1/12}$ (for TBC with $n_f=1$) and $x^{-1/4}$ 
(for TBC with $n_f=3$), in the interval 
$x\in[\lambda_{\rm min}, \lambda_{\rm max}]$,
where we estimate the lower and upper bounds on the eigenvalue spectrum 
of $M^{\dag}M$ from the free-field values, 
$\lambda_{\rm min}=(2m_0a)^{2}$ and 
$\lambda_{\rm max}=64+(2m_0a)^{2}$, respectively.

\begin{figure*}
\begin{center}
\begin{minipage}{3.5in}
\includegraphics[angle=-90,width=3.5in]{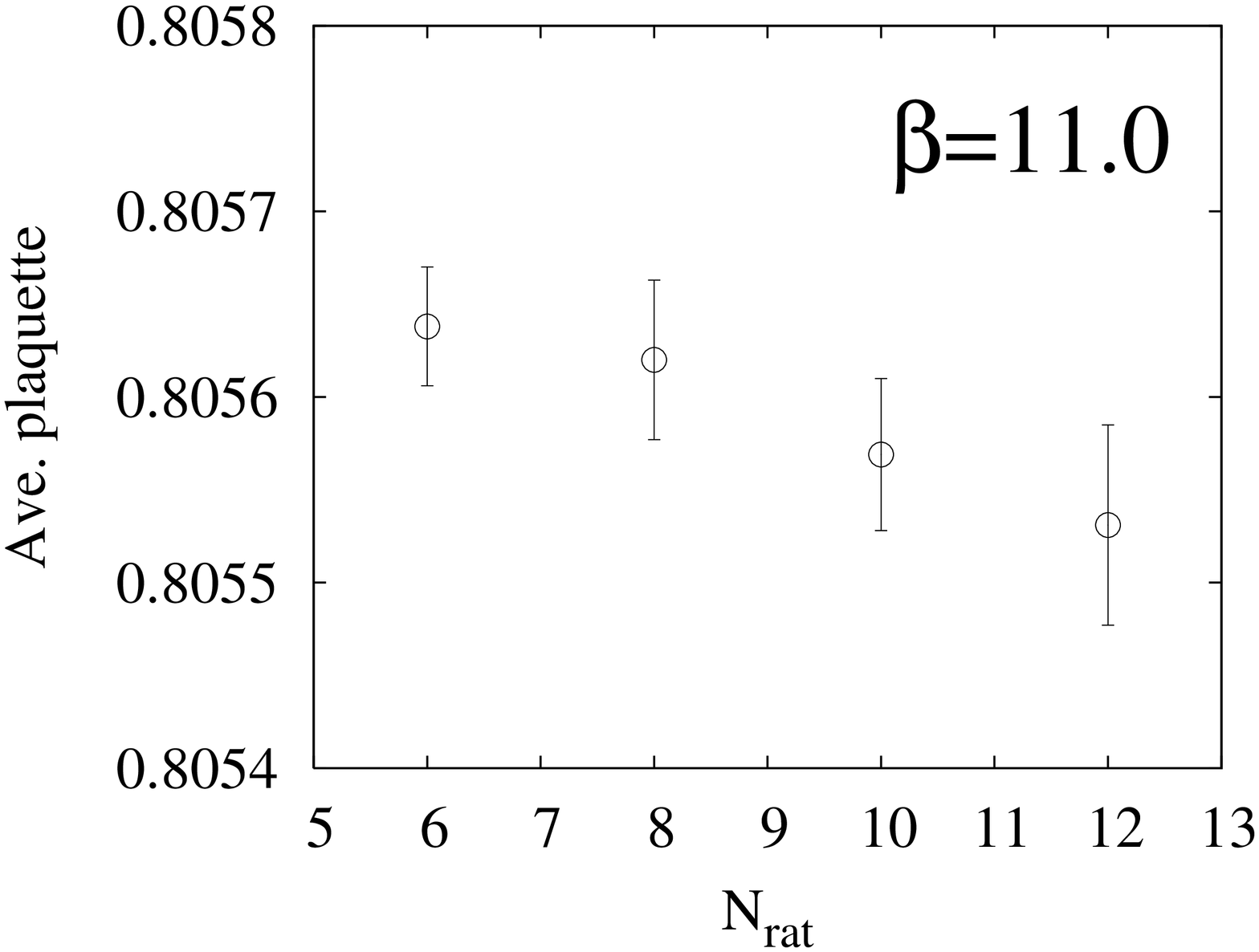}
\end{minipage}
\begin{minipage}{3.5in}
\includegraphics[angle=-90,width=3.5in]{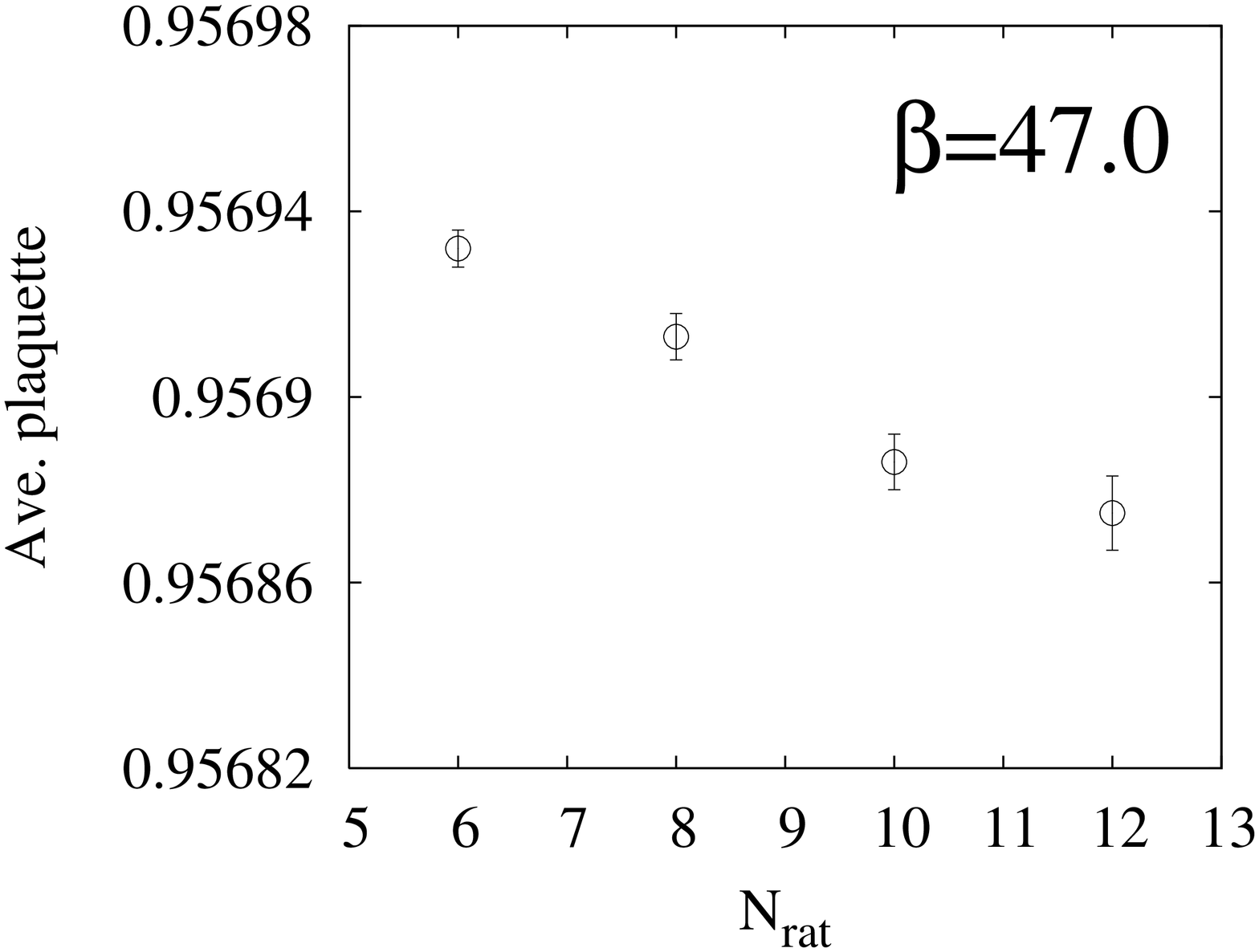}
\end{minipage}
\caption{Average plaquettes computed using different degrees
$N_{\rm rat}$ of the polynomials in the rational approximation
to the unimproved staggered-quark force term. Results are shown for 
$n_f=1$ at $\beta=11$ and 47.}
\label{fig:Nrat}
\end{center}
\end{figure*}

Results for degrees $N_{\rm rat}=6$, 8, 10 and 12 are shown in 
Fig.~\ref{fig:Nrat}, for two values of the bare coupling. The
maximum error in the rational approximation, over the requisite
eigenvalue range, falls below about $10^{-5}$ at around $N_{\rm rat}=10$,
and it appears as well that at larger orders there is little change 
in the ensemble averages of the plaquette, within the statistical 
errors. We therefore use
$N_{\rm rat}=10$ throughout the rest of this study (on the other hand, 
only a 5\% difference in performance is observed for each additional 
order in the approximation). The RHMC code is found to be
about two-times slower compared to the $R$ algorithm, using
the same $\Delta t$ and number of molecular-dynamics steps. 
Given that the acceptance rate varies from about 90\% at $\beta=11$ to 
about 50\% at $\beta=47$ (see Table~\ref{UnimpParams}), we conclude 
that the RHMC algorithm is about 2--4 times more expensive;
however, we generated four sets of ensembles at different step-sizes
for the $R$ algorithm, in order to extrapolate to $\Delta t=0$,
hence we find that the two algorithms are, in practice, of
comparable cost for unimproved-staggered quarks. 

\subsection{\label{S:OtherSystematics}Other algorithmic systematics}

\begin{figure*}
\centering
\begin{minipage}{3.4in}
\includegraphics[angle=-90,width=3.4in]{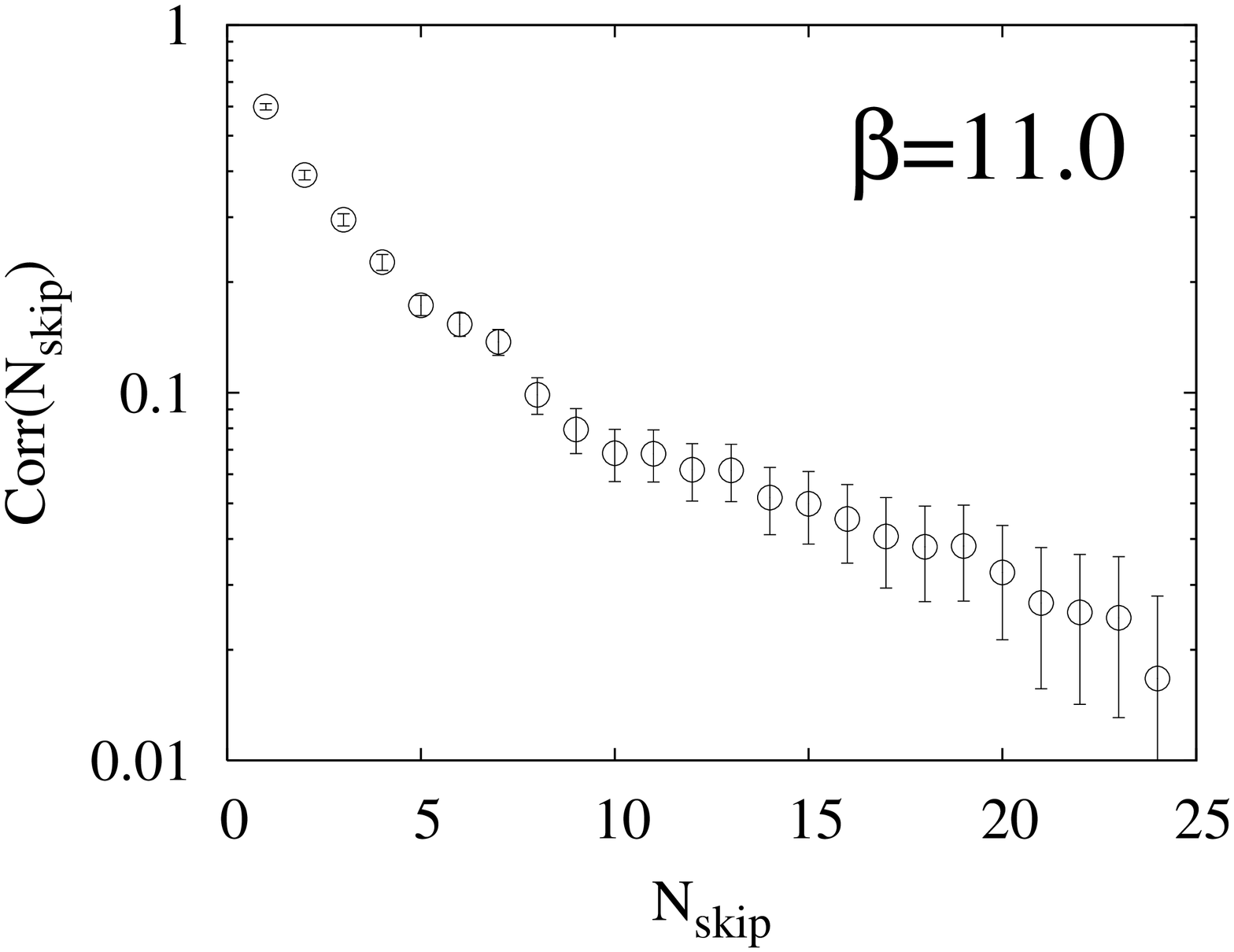}
\end{minipage}
\begin{minipage}{3.4in}
\includegraphics[angle=-90,width=3.4in]{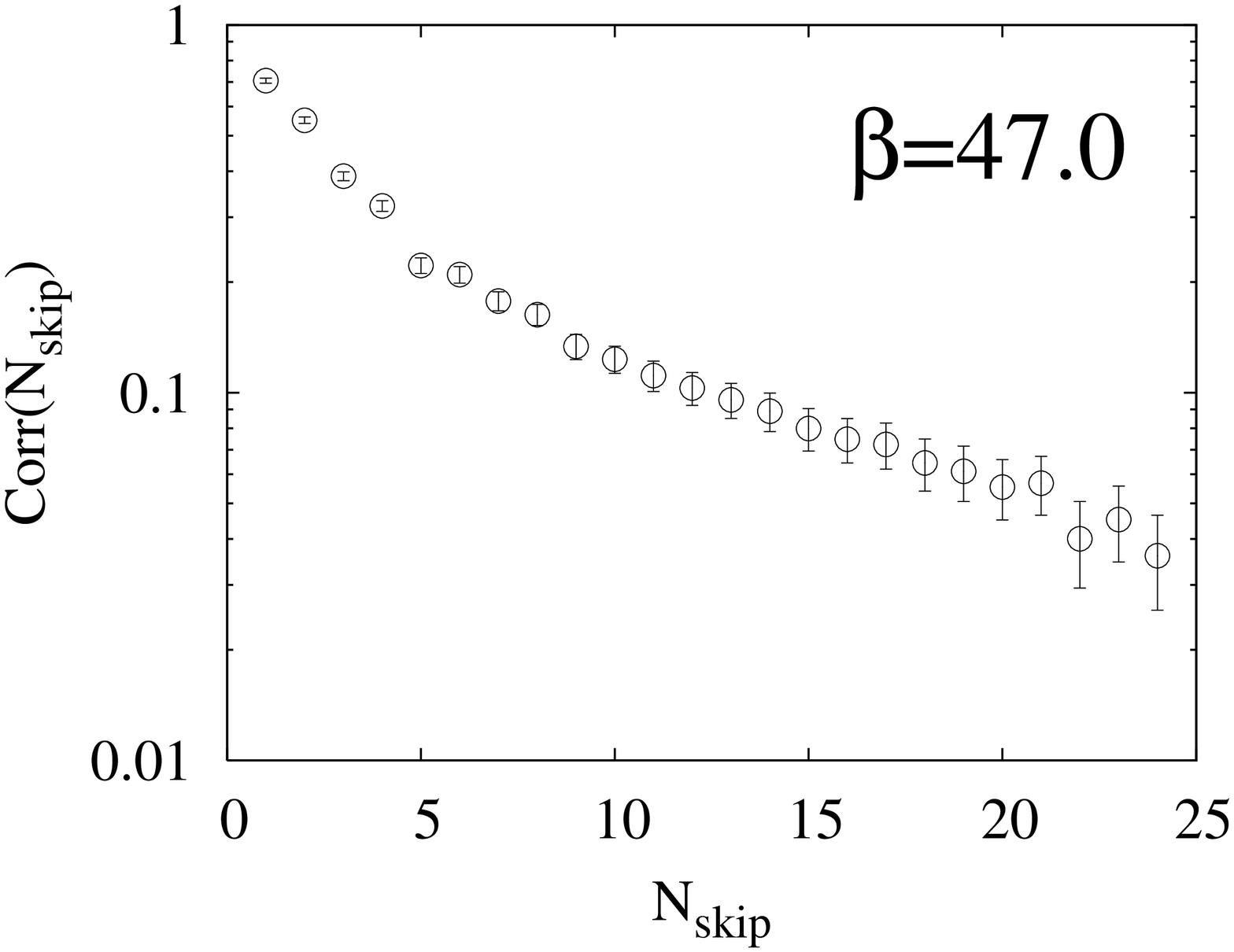}
\end{minipage}
\caption{Autocorrelation plots for the $3 \times 3$ loops for the ``asqtad'' 
actions, at two different couplings. Configurations were generated by
the $R$ algorithm, with $\Delta t=0.01$ and 50 molecular-dynamics
steps per trajectory.}
\label{fig:tau}
\end{figure*}

\begin{figure*}
\centering
\begin{minipage}{3.5in}
\includegraphics[angle=-90,width=3.5in]{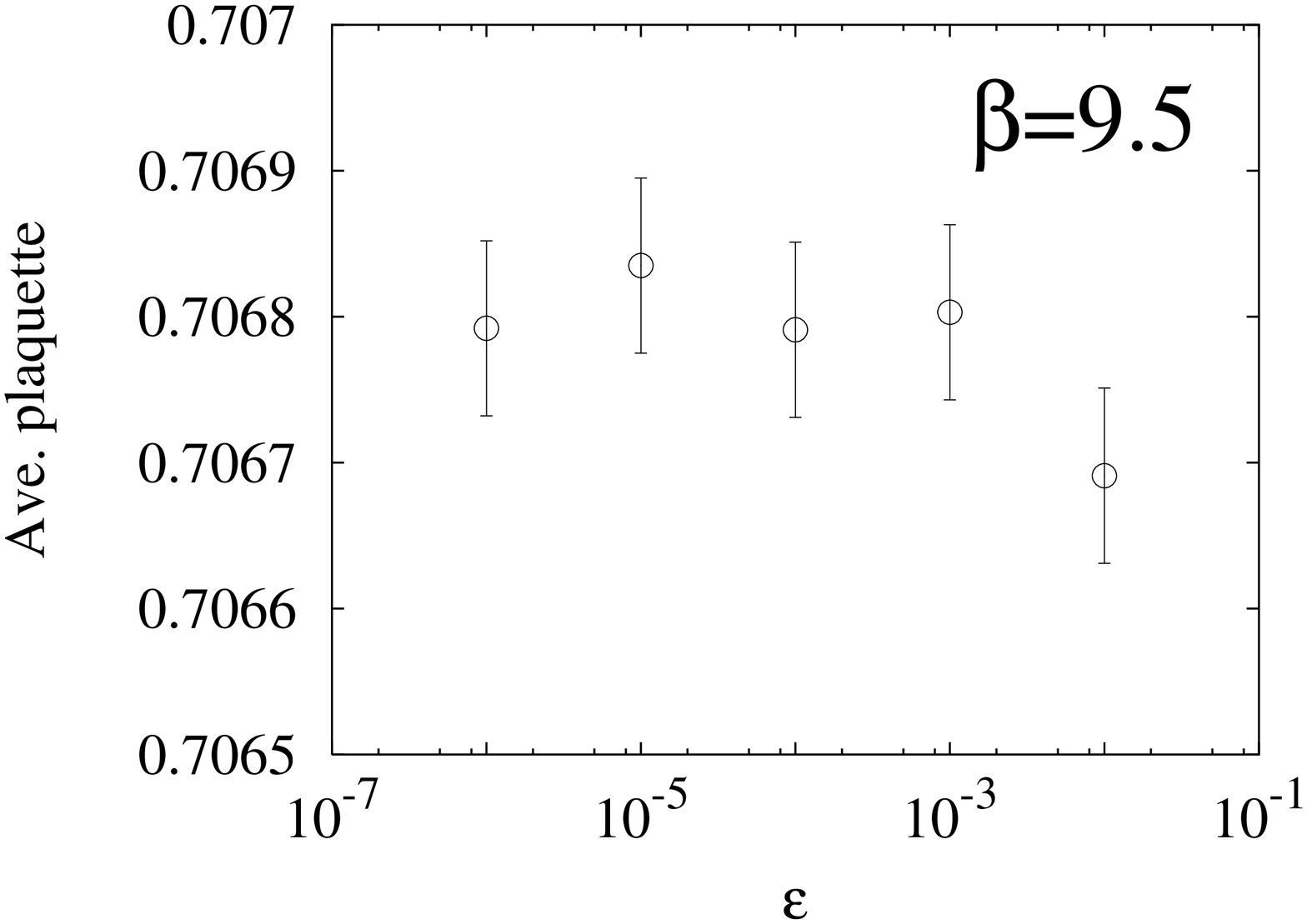}
\end{minipage}
\begin{minipage}{3.5in}
\includegraphics[angle=-90,width=3.5in]{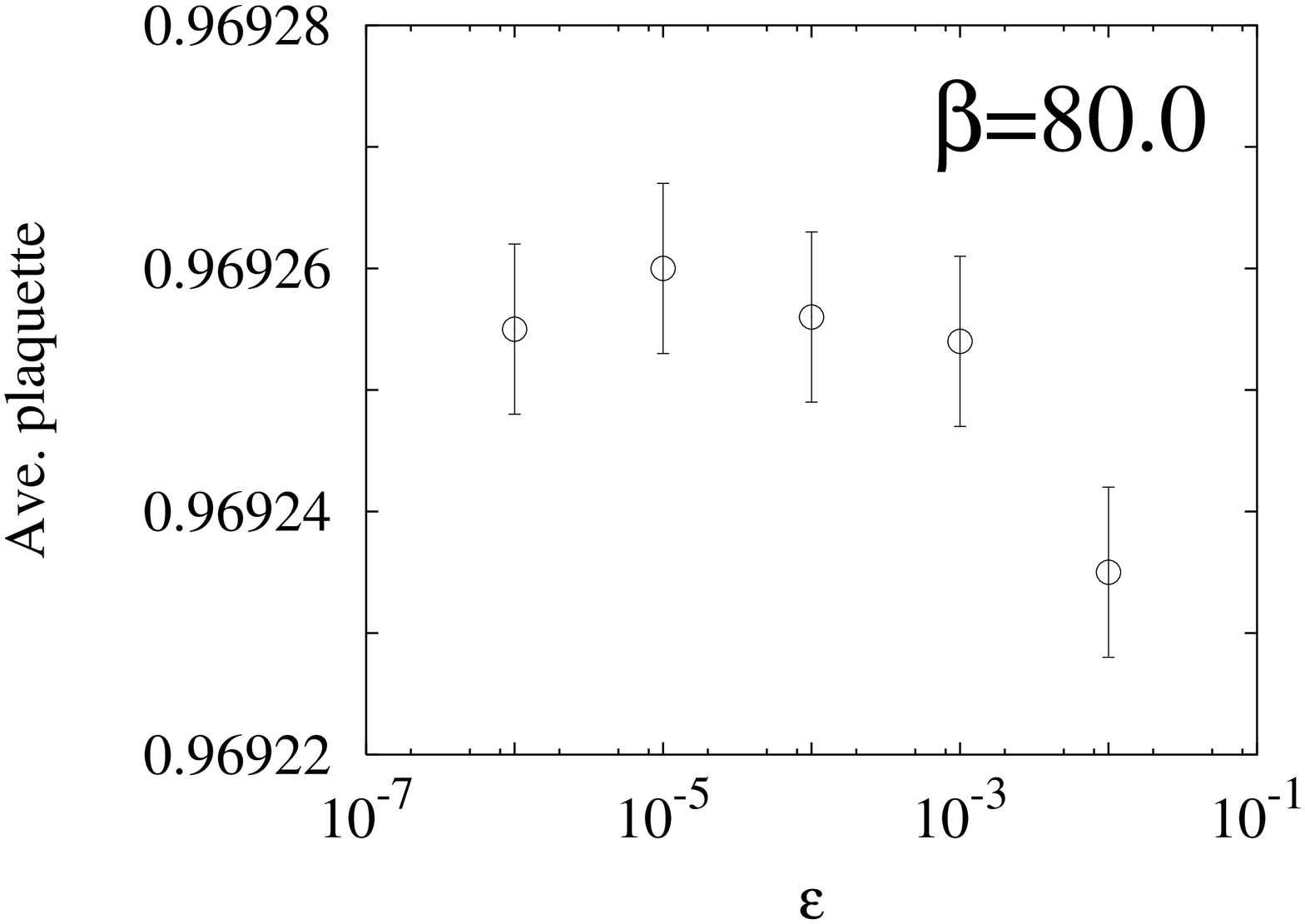}
\end{minipage}
\caption{Average plaquettes computed with different convergence
criteria for the matrix inversion by the stabilized
bi-conjugate-gradient method. Results are shown for the ``asqtad'' actions 
at two couplings.}
\label{fig:cgtest}
\end{figure*}

We have explicitly computed autocorrelation times for the simulation 
data. This is important because gauge-field fluctuations are
suppressed at weak couplings, which should lead to longer autocorrelation
times as the coupling is decreased; moreover, little information on 
autocorrelations in the weak-coupling phase is available from other 
studies. The autocorrelation for a set of measurements 
$\mathcal{O}_{i}$ of some observable is defined, as usual, by
\begin{equation}
   \textrm{Corr}(N_{\rm skip})
   = \frac{\sum_i (\mathcal{O}_{i}-\bar{\mathcal{O}})
                  (\mathcal{O}_{i+N_{\rm skip}}-\bar{\mathcal{O}}) }
   {\sum_i (\mathcal{O}_{i}-\bar{\mathcal{O}})^{2}} ,
\end{equation}
where $N_{\rm skip}$ is the number of ``skipped'' trajectories,
and where the autocorrelation time $\tau$ is estimated from 
$\textrm{Corr}(N_{\rm skip}) \sim e^{-N_{\rm skip}/\tau}$.
The autocorrelation time is observable-dependent, and one generally 
expects to have longer autocorrelation times for a ``larger'' observable, 
such as a larger Wilson loop. Figure~\ref{fig:tau} shows the 
autocorrelation function for the $3\times 3$ Wilson loop, the largest 
considered here, at two couplings, for the ``asqtad'' actions. Note that 
the autocorrelation time is longer for the larger value of $\beta$, as 
expected. Based on these results, we have chosen to skip 20 trajectories 
between measurements for the ``asqtad'' simulations. In the case of the 
RHMC algorithm that was used for the unimproved action, where the 
lowest acceptance rate is about 50\%, 40 trajectories were 
skipped between measurements.

Another source of systematic error is the accuracy of the matrix inversion,
which affects both the $R$ algorithm and the RHMC algorithm. The Monte Carlo 
accept/reject step in the RHMC algorithm cannot remove this error since 
matrix inversions are also required in computing the action for this setp.
Matrix inversions $(M^{\dag}M)^{-1}\phi=x$ were done by the stabilized
bi-conjugate-gradient method, with a convergence criterion
$||(M^{\dag}M)x-\phi||<\epsilon$. We tested the precision of
the inversion by comparing ensemble averages computed with different
values of $\epsilon$. Results are shown in Fig.~\ref{fig:cgtest}
for the average plaquette for the ``asqtad'' actions. Results are
consistent within statistical errors for $\epsilon \alt 10^{-3}$;
we used $\epsilon=10^{-5}$ in the production runs.

An additional systematic error arises from propagation 
of the uncertainties in the couplings $\alpha_V(q_{11}^*)$, which are 
due to the statistical errors in the measured values of the 
$1\times1$ plaquette, from which the couplings are extracted.
The errors in $\alpha_V(q_{11}^*)$ propagate through
to the couplings $\alpha_V(q_{RT}^*)$ that are used in the fits to the
larger Wilson loops. This could be accounted for by including the 
associated uncertainty in the scale parameters $\Lambda_V$ 
(cf. (Eq.~\ref{LambdaV})), in the augmented $\chi^2$ that is used in the 
Bayesian analysis (see Sect.~\ref{S:Fitting} below). We have instead 
propagated the error in the coupling by using a first-order approximation 
to the perturbative expansions (which is adequate for this purpose),
$-\ln W_{RT}/ 2(R+T) \approx c_{1;RT} \, \alpha_{V}(q_{RT}^*)$.
This implies that the statistical uncertainty 
$\Delta[\ln W_{11}]_{\rm MC}$ in the $1\times 1$ loop ``induces'' an 
additional uncertainty $\Delta[\ln W_{RT}]_{\rm induced}$ in the
$R\times T$ loop, beyond its own statistical error, through the
coupling, given by
\begin{equation}
   \Delta[\ln W_{RT}]_{\rm induced} \, \approx \,
   \frac{\ln W_{RT}}{\ln W_{11}} \times \Delta[\ln W_{11}]_{\rm MC} \, .
\end{equation}
It turns out that the simulation error in $\ln W_{RT}$ grows
more rapidly than $-\ln W_{RT}$ itself, as the loop
gets larger, hence the ``induced'' error becomes less important for
larger loops; a representative comparison of the errors for various loops
is shown in Table~\ref{InducedErrors}. The sum of the two correlated 
errors (statistical and ``induced'') is used in the fits for each
Wilson loop.

\begin{table}
\centering
\begin{tabular}{|c||c|c|c|}
\hline
Loop &
$-\langle\ln W_{RT}\rangle$ &
$\Delta[\ln W_{RT}]_{\rm MC}$ &
$\Delta[\ln W_{RT}]_{\rm induced}$ \\
  &  & $(\times 10^{-5}$) & $(\times 10^{-5}$) \\
\hline
1$\times$1 & 0.13976 & 2.2  &  -- \\
1$\times$2 & 0.24382 & 4.6  & 3.8  \\
1$\times$3 & 0.34158 & 7.9  & 5.3  \\
2$\times$2 & 0.39270 & 8.8  & 6.1  \\
2$\times$3 & 0.52235 & 13.7 & 8.1  \\
3$\times$3 & 0.66857 & 20.2 & 10.4 \\
\hline
\end{tabular}
\caption{Comparison of simulation error and ``induced'' error (due
to the uncertainty in the coupling in the perturbative expansion),
for various Wilson loops, for the unimproved actions at 
$\beta=16$ with $n_f=1$.}
\label{InducedErrors}
\end{table}

\subsection{\label{S:Fitting}Fitting and truncation errors}

\begin{table}
\begin{center}
\begin{tabular}{|c|c|c|c|}
\hline
\multicolumn{4}{|c|}{{\bf Dependence on $N$
($\bar\sigma=1.5$)}}
\\ \hline
$c_{n}$ & $N=4$     & $N=5$      & $N=6$      \\
\hline
$c_{1}$ & 1.4334(6) & 1.4334(6) &  1.4334(6) \\
$c_{2}$ & -1.37(4)  & -1.37(4)  & -1.37(4)   \\
$c_{3}$ & -0.91(56) & -0.91(56) & -0.91(56)  \\
$c_{4}$ & -0.1(15)  & -0.1(15)  & -0.1(15)   \\
$c_{5}$ & --        & 0.0(15)   & 0.0(15)    \\
$c_{6}$ & --        & --        & 0.0(15)    \\
\hline
$\chi^2_{\rm aug}$/dof & 0.56   & 0.56      & 0.56       \\
\hline
\end{tabular}
\end{center}
\begin{center}
\begin{tabular}{|c|c|c|c|c|}
\hline
\multicolumn{5}{|c|}{{\bf Dependence on $\bar\sigma$ ($N=6$)}}
\\ \hline
$c_{n}$ & $\bar\sigma=0.5$ & $\bar\sigma=1.0$ & $\bar\sigma=1.5$ & $\bar\sigma=5.0$ \\
\hline
$c_{1}$    & 1.4338(4)    & 1.4335(5) & 1.4334(6) & 1.4333(6)  \\
$c_{2}$    & -1.40(3)     & -1.38(4)  & -1.37(4)  & -1.36(4)   \\
$c_{3}$    & -0.49(38)    & -0.79(51) & -0.91(56) & -0.98(77)  \\
$c_{4}$    & 0.0(5)      & -0.1(10)  & 0.0(15)   & -0.5(50)   \\
$c_{5}$    & 0.0(5)      & 0.0(10)   & 0.0(15)   & 0.0(50)    \\
$c_{6}$    & 0.0(5)      & 0.0(10)   & 0.0(15)   & 0.0(50)    \\
\hline
$\chi^2_{\rm aug}$/dof  & 2.8     & 0.93      & 0.56      & 0.28    \\
\hline
\end{tabular}
\caption{Fit results for the $2\times2$ Wilson loop, for the 
``asqtad'' actions with $n_f=1$. The upper table shows the dependence 
of the results on the number of terms $N$ in the fit, for a fixed prior 
width $\bar\sigma=1.5$, while the lower table shows the dependence on 
$\bar\sigma$ for fixed $N=6$. The augmented $\chi^2$ per degree-of-freedom 
(dof) is also shown. The central values for the prior coefficients are 
$\bar c_n=0$. Diagrammatic perturbation theory yields $c_1=1.4339(0)$,
$c_2=-1.400(2)$ and $c_3=-0.52(7)$ \cite{HQalphaPT}. More accurate 
Monte Carlo estimates of $c_2$ and $c_3$ are obtained in 
Sect.~\ref{S:Results}, using diagrammatic perturbation theory 
to constrain the lower-order terms.}
\label{UnimpPriors}
\end{center}
\end{table}

We use constrained curve fitting \cite{LepageFit} in our fits
to  Eq.~(\ref{lnWRT}), which allows one to incorporate the assumption 
of a convergent perturbation series in a natural way; in particular, 
a large number of higher-order terms can be included in the fit, without 
spoiling the quality of the results for lower-order terms that 
can be resolved by the data.

Constrained curve fitting is motivated by Bayesian analysis; we refer 
the reader to Ref.~\cite{LepageFit} for an overview. In practice, we minimize 
an ``augmented'' least-squares fit function $\chi^2_{\rm aug}$, 
given by
\begin{equation}
   \chi^2_{\rm aug}
   \equiv \chi^2+\sum_{n=1}^N 
   \frac{(c_n - \bar c_n)^2}{\bar\sigma_n^2},
\label{Bayes}
\end{equation}
where $\chi^2$ is the usual weighted sum-of-squared errors in 
fit to the Monte Carlo data, and $N$ is the number of terms in 
the fit (cf.\ Eq.~(\ref{lnWRT})),
which we take to be larger than the order that we anticipate can
be resolved by the data. 
Fitting the data with $\chi^2_{\rm aug}$ favors $c_n$'s in the interval 
$\bar c_n \pm \bar\sigma_n$, which are known collectively as ``priors.''

The values of $\bar c_n$ and $\bar \sigma_n$ are to be chosen based on 
theoretical expectations. In the present case one expects that $c_n = O(1)$, 
if the perturbative expansion is convergent. 
We therefore take $\bar c_n=0$, and use a single prior ``width'' for 
all orders, $\bar\sigma_n \equiv \bar\sigma$, which should be of $O(1)$.
The optimal value for a prior such as $\bar\sigma$ can sometimes
be determined from the data itself, by maximizing the probability of 
obtaining the data, given the prior information, as a function of the 
prior itself; this is the so-called ``empirical'' 
Bayes method (for details see Ref.~\cite{LepageFit}). 

The sensitivity of the data to a given order in the expansion
is reflected in the fit results; coefficients that are well determined by 
the data are relatively insensitive to changes in $\bar\sigma$
and in the number of terms $N$ in the fit, while fit results
for coefficients that are poorly or not at all constrained by the
data simply reproduce the priors. We expect that the 
statistical quality of the simulation data here should allow for 
useful determinations of the three leading orders in the 
perturbative expansion.

To illustrate the quality of the constrained fits, some representative 
results are given in Table~\ref{UnimpPriors}, here for the $2\times2$
Wilson loop for the ``asqtad'' actions with $n_f=1$.
The effects of varying the number of terms $N$ in the fit, and of varying 
the prior width $\bar\sigma$, are shown. We see that the results for the 
three leading orders in the perturbative expansion, $c_1, c_2$, and $c_3$,
are very insensitive to the details of the priors, while the fit  
returns the prior information for the fourth- and higher-order terms,
which simply indicates that the data are not accurate enough to resolve 
those terms, as anticipated.

The results for the three leading orders are in excellent agreement with
diagrammatic perturbation theory, within the fit errors,
of a few parts in $10^4$ for $c_1$, and a few percent for $c_2$;
a reasonable NNLO signal $c_3 \approx -1$ is also obtained, 
which is remarkable, given
how little {\it a priori\/} information went into the fit
(more accurate results for $c_2$ and $c_3$ are obtained in 
Sect.~\ref{S:Results}, using diagrammatic perturbation theory 
to constrain the lower orders).

The results for the $\chi^2_{\rm aug}$ of the fits also suggest that an
``optimal'' prior width can be determined from the data, with
$\bar\sigma \sim 1.0$--1.5, which we also find using the 
optimization procedure given by the ``empirical'' Bayes method, alluded
to above. Hence the simulation data are indeed consistent with the 
expectation that perturbation theory is reliable. For the rest
of the fits in this paper we use $N=6$, $\bar c_n = 0$, and
$\bar\sigma=1.5$.

\begin{figure*}
\begin{center}
\begin{minipage}{2.25in}
\includegraphics[angle=-90,width=2.25in]{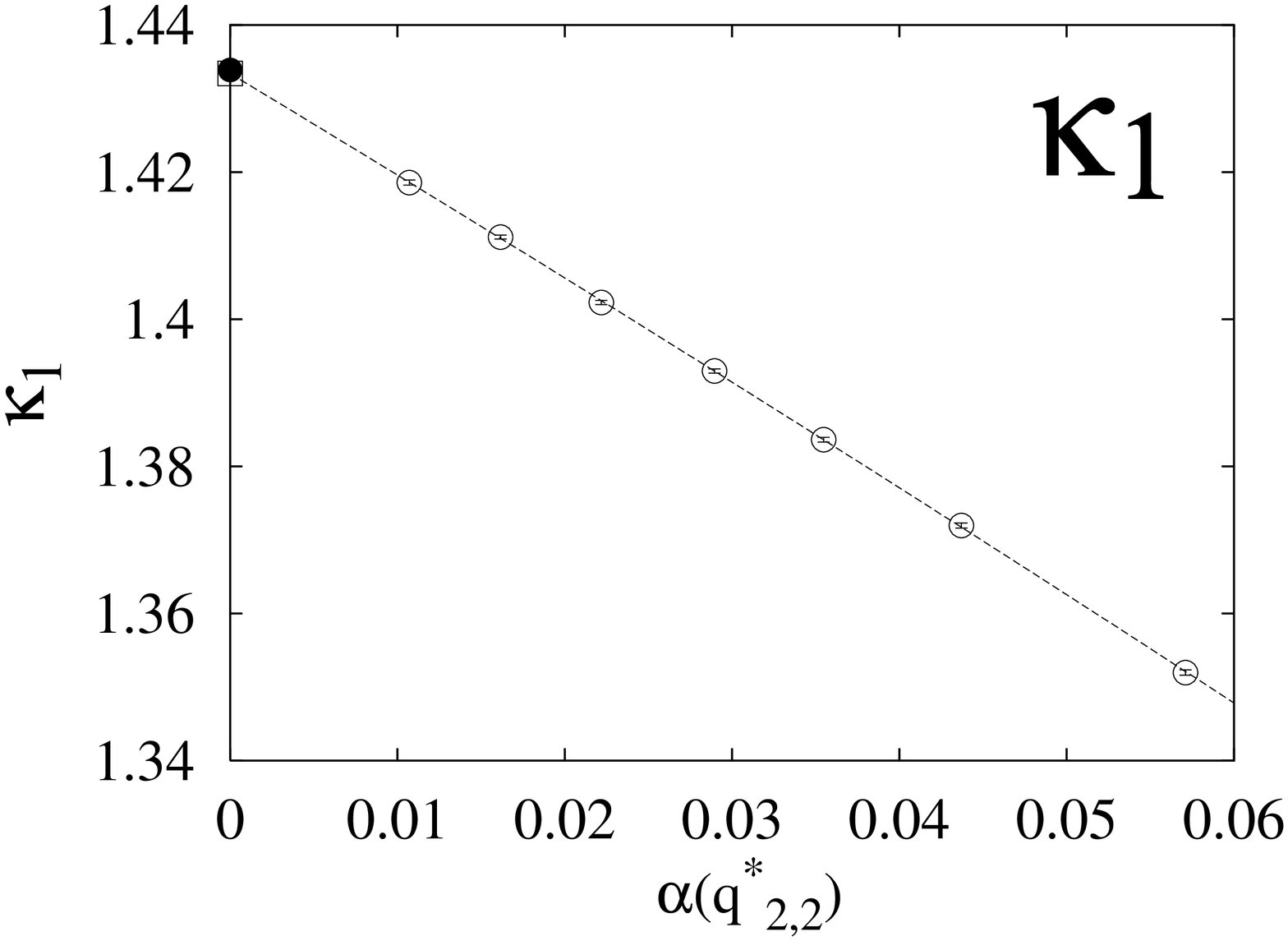}
\end{minipage}
\begin{minipage}{2.25in}
\includegraphics[angle=-90,width=2.25in]{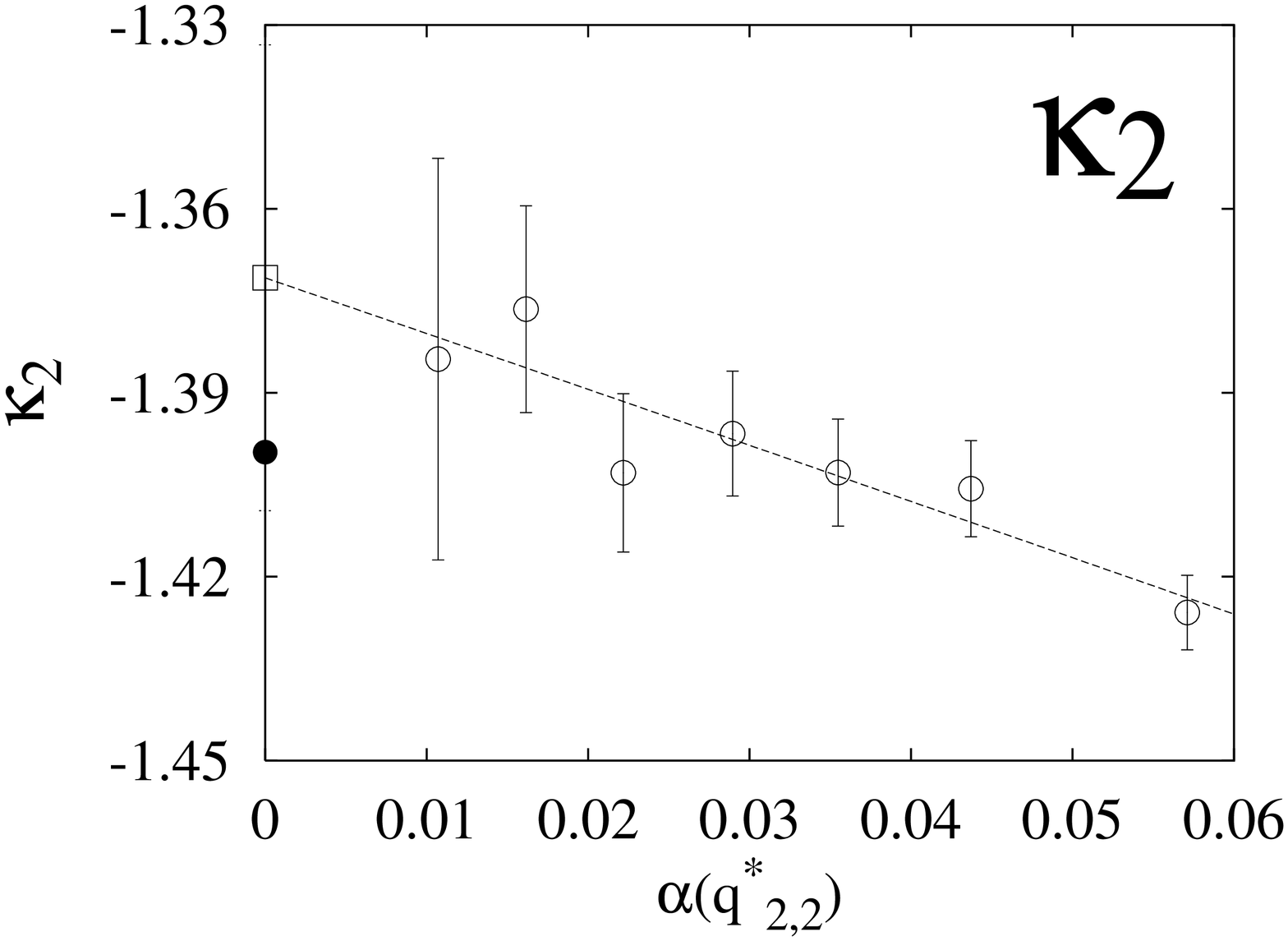}
\end{minipage}
\begin{minipage}{2.25in}
\includegraphics[angle=-90,width=2.25in]{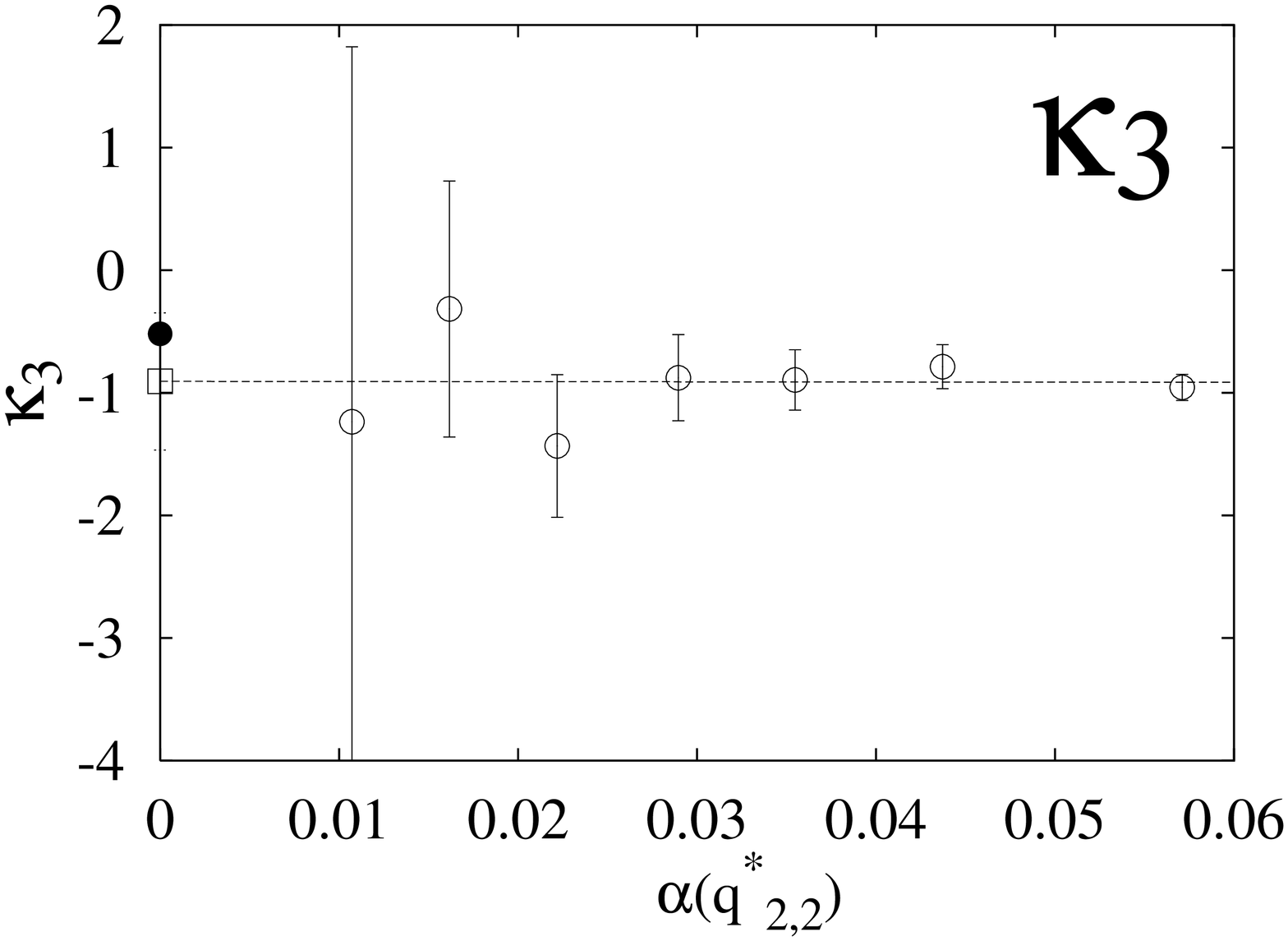}
\end{minipage}
\caption{Plots of the simulation results for 
$\kappa_1$, $\kappa_2$, and $\kappa_3$, for the $2\times 2$ 
Wilson loop, for the unimproved actions with $n_f=1$. 
The simulation parameters are given in Table~\ref{UnimpParams}.
The intercepts marked by open squares ($\Box$) in the plot for 
each $\kappa_n$ give the results of the fits to the corresponding $c_n$, 
while the diagrammatic values \cite{HQalphaPT} are indicated by 
filled circles ($\bullet$).}
\label{Unimpkappas}
\end{center}
\end{figure*}

\begin{table*}
\begin{center}
\begin{tabular}{|c|c|c|c||c|c|c|}
\hline
\multicolumn{7}{|c|}{Number of flavours: $n_f=1$} \\
\multicolumn{1}{|c}{\ } & 
\multicolumn{3}{c}{Monte Carlo method} & 
\multicolumn{3}{c|}{Diagrammatic values} \\ 
\hline
Loop & $c_{1}$ & $c_{2}$ & $c_{3}$ & $c_{1}$ & $c_{2}$ & $c_{3}$ \\ 
\hline
1$\times$2 & 1.2038(4) & -1.327(29) & -1.26(46) 
           & 1.2039(0) & -1.335(0)  & -1.10(3)   \\
1$\times$3 & 1.2586(5) & -1.253(36) & -1.43(56)   
           & 1.2589(0) & -1.277(1)  & -0.95(6)   \\
2$\times$2 & 1.4334(6) & -1.368(39) & -0.95(59)
           & 1.4339(0) & -1.400(2)  & -0.52(8)   \\
2$\times$3 & 1.5163(7) & -1.281(47) & -1.11(71) 
           & 1.5172(0) & -1.351(3)  & -0.26(12)  \\
3$\times$3 & 1.6080(8) & -1.230(56) & -0.45(82)  
           & 1.6090(0) & -1.298(6)  & 0.67(25)   \\ 
\hline
\end{tabular}
\end{center}
\begin{center}
\begin{tabular}{|c|c|c|c||c|c|c|}
\hline
\multicolumn{7}{|c|}{Number of flavours: $n_f=3$} \\
\multicolumn{1}{|c}{\ } & 
\multicolumn{3}{c}{Monte Carlo method} & 
\multicolumn{3}{c|}{Diagrammatic values} \\ 
\hline
Loop & $c_{1}$ & $c_{2}$ & $c_{3}$ & $c_{1}$ & $c_{2}$ & $c_{3}$
\\ \hline
1$\times$2 & 1.2039(4) & -1.480(28) & -0.28(45)
           & 1.2039(0) & -1.485(0)  & -0.11(9)   \\
1$\times$3 & 1.2590(5) & -1.444(34) & -0.06(54)  
           & 1.2589(0) & -1.437(1)  & 0.02(11)   \\
2$\times$2 & 1.4337(5) & -1.529(38) & 0.02(60)  
           & 1.4339(0) & -1.551(2)  & 0.51(13)   \\
2$\times$3 & 1.5169(6) & -1.476(46) & 0.20(71)  
           & 1.5172(0) & -1.513(4)  & 0.73(16)   \\
3$\times$3 & 1.6088(8) & -1.431(53) & 0.65(81)  
           & 1.6090(0) & -1.463(11) & 1.62(30)   \\ 
\hline
\end{tabular}
\caption{Perturbative coefficients of various small Wilson loops 
for the unimproved actions, with $n_f=1$, and with $n_f=3$. 
The coefficients from diagrammatic perturbation theory \cite{HQalphaPT}
are compared with the results obtained from the Monte Carlo simulations.
These fits to the Monte Carlo data used no diagrammatic input, except in
the determination of the couplings for the expansion Eq.~(\ref{lnWRT}).}
\label{Unimpc123}
\end{center}
\end{table*}

\begin{table*}[!]
\begin{center}
\begin{tabular}{|c|c|c||c|c|}
\hline
\multicolumn{5}{|c|}{Diagrammatic input: $c_1$} \\
\multicolumn{1}{|c}{\ } & 
\multicolumn{2}{c}{$n_f=1$} &
\multicolumn{2}{c|}{$n_f=3$} \\
\hline
Loop & $c_{2}$ & $c_{3}$ & $c_{2}$ & $c_{3}$ \\ 
\hline
1$\times$2 & -1.332(9)  & -1.19(22) & -1.480(10) & -0.28(25)\\
1$\times$3 & -1.270(11) & -1.20(27) & -1.434(12) & -0.19(30)\\
2$\times$2 & -1.403(12) & -0.48(29) & -1.541(13) & 0.20(33) \\
2$\times$3 & -1.342(14) & -0.28(34) & -1.499(16) & 0.52(39) \\
3$\times$3 & -1.292(17) & 0.38(41)  & -1.442(20) & 0.80(46) \\ 
\hline
\end{tabular}
\hspace{0.5cm}
\begin{tabular}{|c|c|c||c|c|}
\hline
\multicolumn{5}{|c|}{Diagrammatic input: $c_{1,2}$} \\
\multicolumn{1}{|c}{\ } & 
\multicolumn{2}{c}{$n_f=1$} &
\multicolumn{2}{c|}{$n_f=3$} \\
\hline
Loop & \multicolumn{2}{c||}{$c_{3}$} & \multicolumn{2}{|c|}{$c_{3}$} 
\\ \hline
1$\times$2 & \multicolumn{2}{c||}{-0.11(9) }& \multicolumn{2}{|c|}{-0.17(10)} \\
1$\times$3 & \multicolumn{2}{c||}{ 0.02(11)}& \multicolumn{2}{|c|}{-0.12(11)} \\
2$\times$2 & \multicolumn{2}{c||}{ 0.51(13)}& \multicolumn{2}{|c|}{ 0.42(13)} \\
2$\times$3 & \multicolumn{2}{c||}{ 0.73(16)}& \multicolumn{2}{|c|}{ 0.81(16)} \\
3$\times$3 & \multicolumn{2}{c||}{ 1.62(30)}& \multicolumn{2}{|c|}{ 1.16(27)} \\ 
\hline
\end{tabular}
\caption{Monte Carlo results for perturbative coefficients for
the unimproved actions, where in the left table the first-order term 
is set to its diagrammatic value, while in the right table both 
$c_1$ and $c_2$ are set to their diagrammatic values.}
\label{Unimpc23}
\end{center}
\end{table*}

\section{\label{S:Results}Results}
\subsection{\label{R:Strategy}Fitting strategy}

Simulation results for the perturbative coefficients of
various Wilson loops are presented for the unimproved and ``asqtad''
actions in the next two subsections. The results are compared
with a recent NNLO analysis using diagrammatic perturbation theory,
which was done in the infinite-volume limit, and 
for zero quark mass, in Ref.~\cite{HQalphaPT}. 

We perform several types of fits for each set of actions.
Fits are first done without input from 
diagrammatic perturbation theory for a given Wilson loop (though we 
always assume the relevant scales $q_{RT}^*$, as well as the coefficients 
for the $1\times1$ loop, in order to extract the numerical values of the 
$\alpha_V$ couplings from the simulation data). We then set the 
first-order coefficients to their values from diagrammatic perturbation 
theory, so as to improve the Monte Carlo estimates of the 
second- and third-order terms,
and then we similarly constrain both the first- and second-order coefficients,
so as to obtain the most accurate estimate possible of the third-order terms.
This provides us with very stringent tests of the Monte Carlo method,
and a very precise cross-check of the difficult NNLO diagrammatic
calculations in Ref.~\cite{HQalphaPT}. 

For fits that use diagrammatic values to constrain 
lower-order coefficients, we computed the leading-order 
Feynman diagram loop integrals with TBC, 
for the $8^4$ volume and quark masses that were used 
in the simulations, in order to consistently account for finite-volume
and finite-mass effects in the simulation data.
However these effects are negligible except for the largest
Wilson loops and at the smallest couplings considered here 
[the leading finite-volume corrections are 
$\sim \alpha_V/\mbox{Vol} = O(10^{-4}) \times \alpha_V$, while the
leading mass-dependent corrections are
$\sim \alpha_V^2 (m_0a)^2 = O(0.01) \times \alpha_V^2$;
see also Ref.~\cite{QuenchedHighBeta} for
an extensive finite-volume analysis in the 
pure-gauge Monte Carlo perturbation theory.] 

We also plot the following three residuals
\begin{align}
   \kappa_1 & \equiv \frac{1}{\alpha_V(q_{RT}^*)}
   \left[\frac{-\ln W_{RT}}{2(R+T)}\right] , \\
   \kappa_2 & \equiv \frac{1}{\alpha_V^2(q_{RT}^*)}
   \left[ \frac{-\ln W_{RT}}{2(R+T)} - c_{1;RT} \alpha_V \right] ,
\end{align}
and
\begin{align}
   \kappa_3 & \equiv \frac{1}{\alpha_V^3(q_{RT}^*)}
   \left[ \frac{-\ln W_{RT}}{2(R+T)} - c_{1;RT} \alpha_V
   - c_{2;RT} \alpha_V^2 \right] ,
\label{kappas} 
\end{align}
which provide useful visualizations of the quality of the data
(diagrammatic perturbation theory is used 
for the coefficients in $\kappa_2$ and $\kappa_3$). We 
plot the residuals versus the appropriate coupling $\alpha_V(q_{RT}^*)$, 
and in the limit of zero coupling one should find $\kappa_n \to c_n$
for the particular Wilson loop. Likewise a visible slope in the residual
$\kappa_n$ at small couplings reveals the sensitivity of the data 
to the next-order coefficient $c_{n+1}$.

\subsection{\label{R:Unimproved}Unimproved actions}

The unimproved actions were simulated for $n_f=1$ and $n_f=3$.
The sensitivity of the data to three leading orders in the
perturbative expansions is apparent in the plots of the three 
residuals $\kappa_1$, $\kappa_2$, and $\kappa_3$, which 
are shown for a representative Wilson loop in Fig.~\ref{Unimpkappas}.
The slope and curvature in the plot for $\kappa_1$, for instance, 
indicate the sensitivity to  $c_2$ and $c_3$, while the fact that the 
plot for $\kappa_3$ has no noticeable slope indicates that the data 
are insensitive to $c_4$, within the statistical errors.

\begin{figure*}
\centering
\begin{minipage}{2.25in}
\includegraphics[angle=-90,width=2.25in]{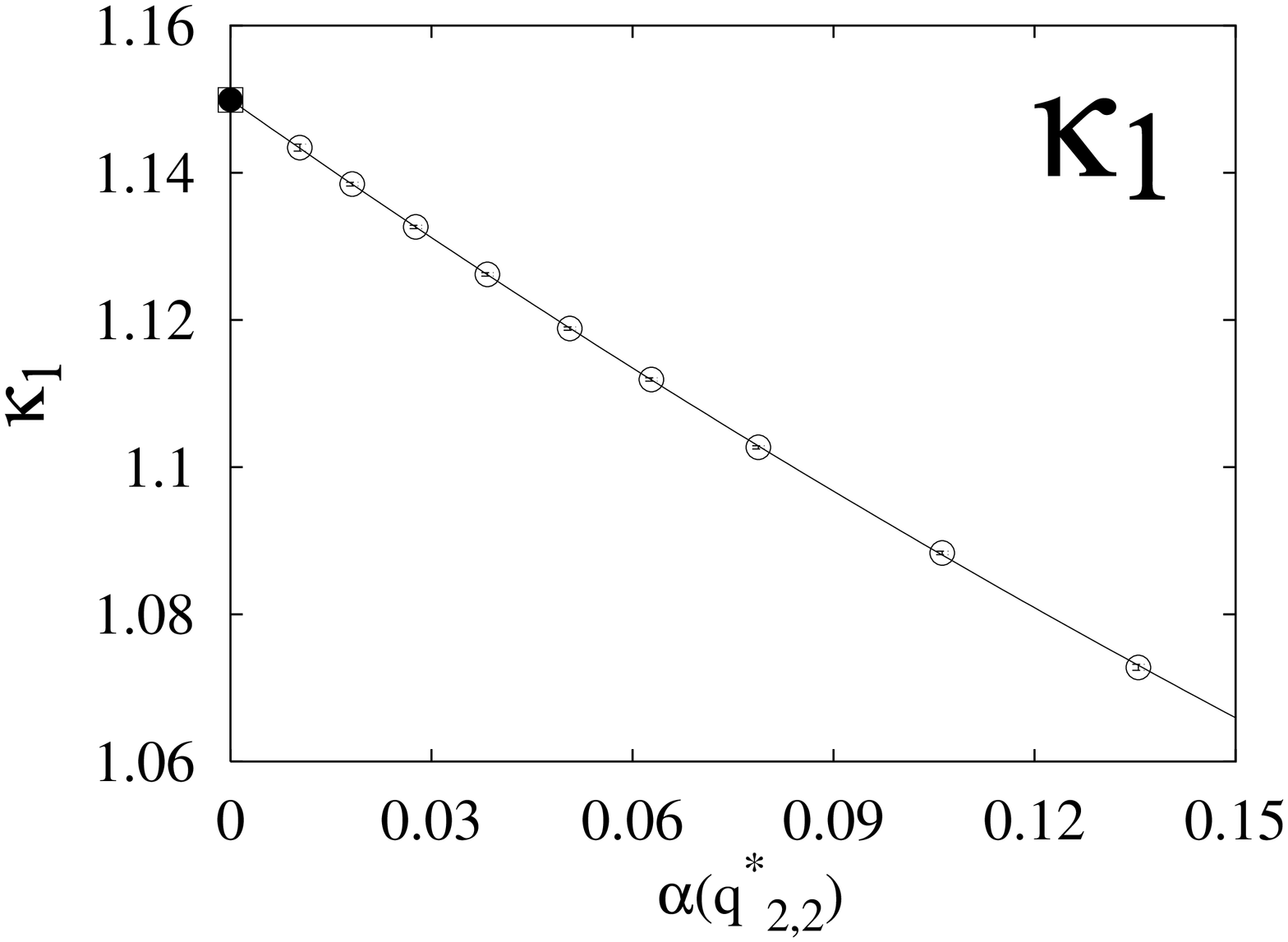}
\end{minipage}
\begin{minipage}{2.25in}
\includegraphics[angle=-90,width=2.25in]{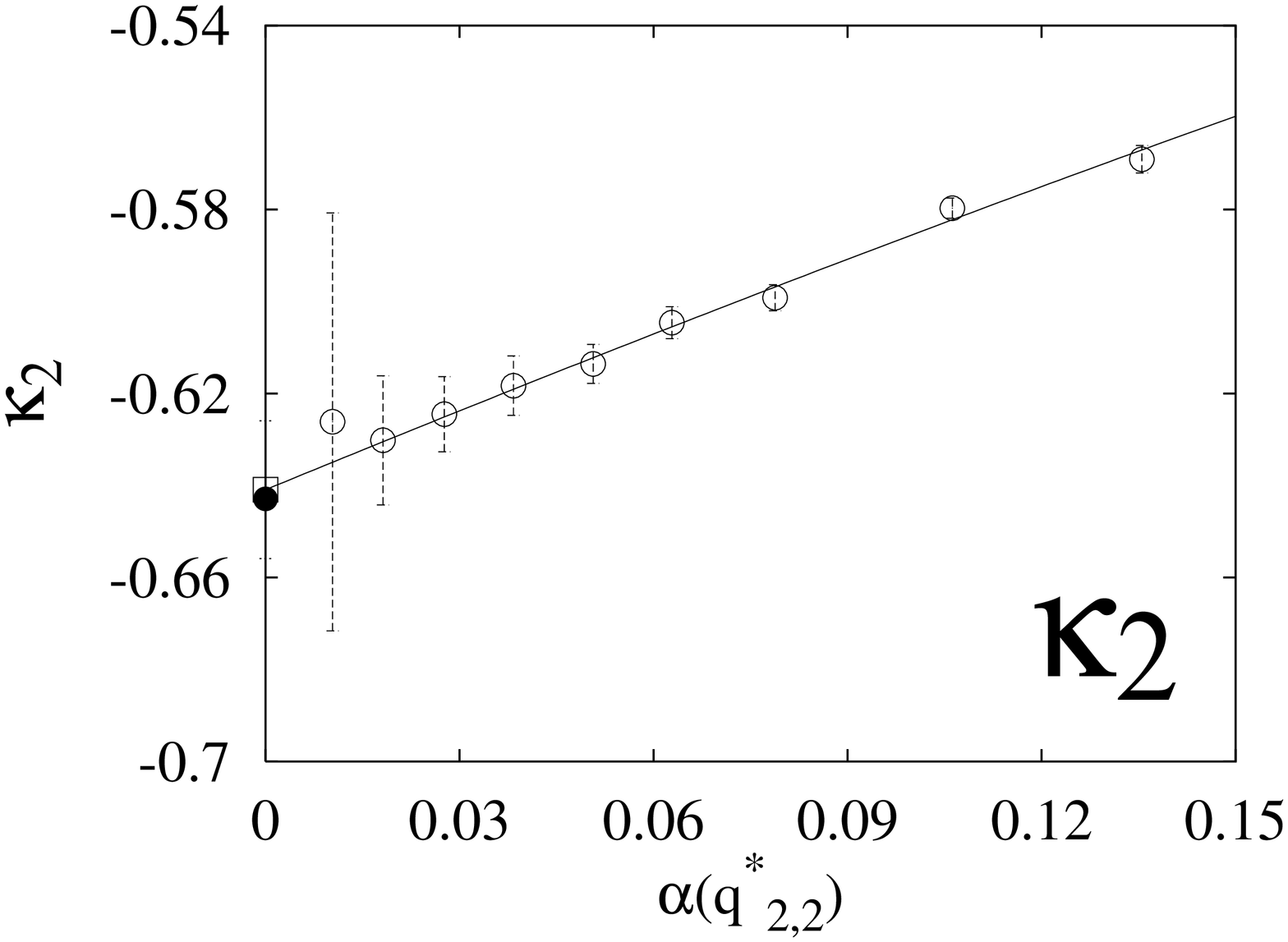}
\end{minipage}
\begin{minipage}{2.25in}
\includegraphics[angle=-90,width=2.25in]{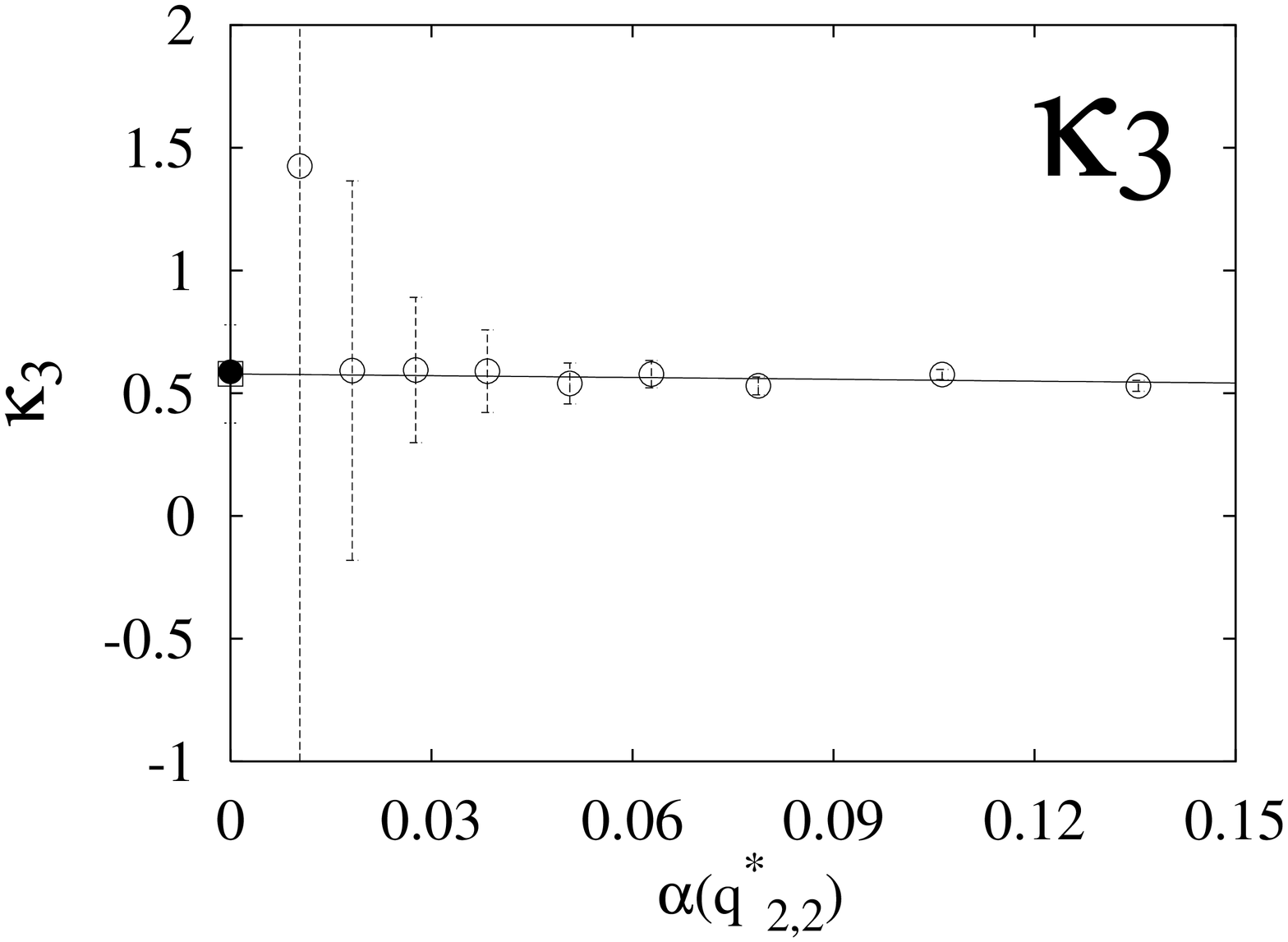}
\end{minipage}
\caption{Plots of $\kappa_1$, $\kappa_2$, and $\kappa_3$
for the 2$\times$2 loop with the ``asqtad'' actions.
Simulation parameters are given in Table~\ref{ImpParams}.
The results of fits to the Monte Carlo data for the
perturbative coefficients are indicated by open squares ($\Box$), 
while the diagrammatic results are shown by filled circles ($\bullet$).}
\label{Impkappas}
\end{figure*}

\begin{table*}
\begin{center}
\begin{tabular}{|c|c|c|c||c|c|c|}
\hline
\multicolumn{1}{|c}{\ } & 
\multicolumn{3}{c}{Monte Carlo method} & 
\multicolumn{3}{c|}{Diagrammatic values} \\ 
\hline
Loop & $c_{1}$ & $c_{2}$ & $c_{3}$ & $c_{1}$ & $c_{2}$ & $c_{3}$
\\ \hline
1$\times$2 & 0.9251(3) & -0.644(13) & 0.20(18) 
           & 0.9252(0) & -0.646(0)  & 0.23(5)    \\
1$\times$3 & 0.9845(3) & -0.599(14) & 0.37(19)  
           & 0.9845(0) & -0.595(1)  & 0.38(6)   \\
2$\times$2 & 1.1499(4) & -0.641(15) & 0.58(20)  
           & 1.1499(0) & -0.643(2)  & 0.59(9)   \\
2$\times$3 & 1.2342(4) & -0.599(19) & 0.88(26) 
           & 1.2341(0) & -0.595(3)  & 0.85(16)  \\ 
3$\times$3 & 1.3235(5) & -0.545(19) & 1.16(23) 
           & 1.3235(0) & -0.522(4)  & 0.96(19)  \\ 
\hline
\end{tabular}
\hspace{0.25cm}
\begin{tabular}{|c|c|c|}
\hline
\multicolumn{3}{|c|}{Diagrammatic input: $c_1$} \\
\hline
Loop & $c_{2}$ & $c_{3}$ \\ 
\hline
1$\times$2 & -0.649(5)  & 0.26(12) \\
1$\times$3 & -0.600(6)  & 0.39(13) \\
2$\times$2 & -0.642(7)  & 0.59(14) \\
2$\times$3 & -0.594(7)  & 0.82(15) \\ 
3$\times$3 & -0.546(8)  & 1.17(17) \\ 
\hline
\end{tabular}
\hspace{0.25cm}
\begin{tabular}{|c|c|}
\hline
\multicolumn{2}{|c|}{Diagrammatic input: $c_{1,2}$} \\
\hline
Loop & $c_{3}$ \\ 
\hline
1$\times$2 & 0.21(4) \\
1$\times$3 & 0.28(5) \\
2$\times$2 & 0.61(6) \\
2$\times$3 & 0.84(8) \\ 
3$\times$3 & 0.84(10) \\ 
\hline
\end{tabular}
\caption{Perturbative coefficients for various small Wilson loops
for the ``asqtad'' actions. There is no diagrammatic input in
the Monte Carlo results in the left-most table (except for the couplings).
In the middle table the Monte Carlo results are shown with
the first-order term set to its diagrammatic value, 
while in the right-most table both $c_1$ and $c_2$ are set to their 
diagrammatic values.}
\label{Impc123c23}
\end{center}
\end{table*}

Fit results for $c_1$, $c_2$, and $c_3$, without input from
diagrammatic perturbation theory (except for the couplings) 
are given in Table~\ref{Unimpc123}, for Wilson loops up to
3$\times$3. As discussed in Sect.~\ref{S:Fitting}, excellent
agreement with diagrammatic calculations is obtained, within
the fit errors, which for $c_1$ are typically a few parts in $10^4$,
and for $c_2$ are typically a few percent. The third order terms
are also resolved, which is remarkable given that no
diagrammatic input for these Wilson loops was used. 
The simulations also clearly resolve the 
perturbative dependence on the number of flavors, which in
the case of $c_2$ shows a change of 3--4 standard deviations
from $n_f=1$ to $n_f=3$.

The accuracy of the Monte Carlo results for the higher-order coefficients 
can be dramatically improved by further constraining the lower-order 
terms using available diagrammatic input. Results are shown in 
Table~\ref{Unimpc23} where we first fix $c_1$ to its diagrammatic value, 
and then fix both $c_1$ and $c_2$; in the first case, the fit errors in 
$c_2$ and $c_3$ are reduced by factors of about 2--3 compared to the 
results in Table~\ref{Unimpc123}, while in the second case the errors 
in $c_3$ are reduced by a factor of about 5. Agreement with the 
diagrammatic values is obtained in all cases, within these greatly 
reduced errors, and in most cases the Monte Carlo results have errors 
that are comparable to those in the diagrammatic evaluations.

\subsection{\label{R:asqtad}The ``asqtad'' actions}

Our results for the perturbative expansions of Wilson loops for
the ``asqtad'' actions are shown in Fig.~\ref{Impkappas}, and
in Table~\ref{Impc123c23}. The data is sensitive 
to the three leading orders of the perturbative expansion, and the 
agreement with diagrammatic results is again impressive. 

We note that the uncertainties in the coefficients obtained from
these simulations are about a factor of 3 smaller than for the 
unimproved results presented in the preceding subsection, which 
owes to the fact that the ``asqtad'' simulations were done at 
larger couplings [the largest coupling for the unimproved
simulations was $\alpha_V(q^*_{11}) \approx 0.06$, while for
the ``asqtad'' simulations it was $\alpha_V(q^*_{11}) \approx 0.13$,
cf.\ Tables~\ref{UnimpParams} and \ref{ImpParams}]. 
This fact should clearly be heeded in future studies, where one
might work at still larger couplings, as long as the theory remains
in the perturbative phase, as judged for example by simulation 
measurements of the Polyakov line, which provides an order
parameter for confinement (a systematic approach to optimizing
the choice of couplings is given in 
Refs.~\cite{QuenchedHighBeta,LepageFit}).

\section{\label{S:Conclusions}Conclusions and Outlook}

Perturbative coefficients of Wilson loops were extracted
from unquenched QCD simulations at weak couplings. Two sets of 
actions were analyzed: unimproved gluon and staggered-quark actions 
in one set, and $O(a^2)$ improved actions in the other. Simulations 
were also done for different numbers of dynamical fermions. 
An extensive analysis of systematic uncertainties was 
made; constrained-curve fitting in particular was used to 
extract as much information as possible from the simulation data. 

The Monte Carlo results for the perturbative coefficients were 
found to be in excellent agreement with calculations using 
diagrammatic perturbation theory, through next-to-next-to leading order. 
Results were obtained for the first-order coefficients with uncertainties 
of a few parts in $10^4$, while the second-order coefficients were obtained 
to a few-percent precision, without any input from diagrammatic
perturbation theory (except for the perturbative expansion of the
$1\times1$ plaquette, which was used to extract the relevant couplings
$\alpha_V(q^*)$ from the simulation data). The results also
show that the Monte Carlo perturbation theory is clearly sensitive to 
the number of dynamical fermion. Furthermore, when the two leading 
perturbative coefficients were constrained to their diagrammatic values, 
the third-order term was obtained with a precision comparable
to that from the NNLO diagrammatic analysis \cite{HQalphaPT},
which required the evaluation of about fifty Feynman diagrams.

These results provide a stringent test of the Monte Carlo method
as applied to highly-improved lattice actions with dynamical fermions, 
and also provide an important high-precision cross-check of the 
perturbation theory input to a recent determination of 
$\alpha_{\overline{\rm MS}}(M_Z)$ by the HPQCD collaboration
\cite{HPQCDalphas}.

Remarkably little computational power was needed, since the simulations
were done on small volumes ($8^4$), thanks to the use of 
twisted boundary conditions, which eliminate lattice zero modes,
and which suppress other finite-volume effects; the entire set
of simulations for the ``asqtad'' actions only required the
equivalent of about 60 months of run-time on a single 3~GHz processor.
Larger lattices might be needed for other quantities, such as quark
propagators \cite{Juge}, that are not as dominated 
by ultraviolet modes as the small Wilson loops analyzed 
here, though an earlier study for pure-gauge 
theories \cite{QuenchedHighBeta} found that 
finite-volume effects in such quantities could be removed by 
working on several relatively small volumes. 

Additional work using this method is in progress, including the 
determination of the NNLO mass renormalization for the NRQCD 
heavy-quark action, and preliminary investigations of 
currents with infrared singularities.

\acknowledgments
We thank Peter Lepage, Christine Davies, and Matthew Nobes for
fruitful conversations. We also thank Peter Lepage for providing us
with his Bayesian-analysis software. We are grateful to
Randy Lewis for providing resources including the use of his 
computer cluster at the University of Regina. Computations were 
also performed on facilities provided by {\tt WestGrid} \cite{WestGrid}.
This work was supported in part by the Natural
Sciences and Engineering Research Council of Canada.



\begin{thebibliography}{99}
\bibitem{PlaqPTNote} We recomputed the leading-order Feynman diagram
loop integrals here for finite-volume ($8^4$) TBC perturbation theory, and 
for the finite quark mass; see Sect.~\ref{R:Strategy}.
\end{thebibliography}
\end{document}